\documentclass[twocolumn, twocolappendix, usenatbib, useAMS]{aastex631}
\usepackage{soul, hyperref, orcidlink}
\usepackage{enumitem, color, graphicx}
\usepackage{fleqn, times, amsmath, amssymb, amsfonts, latexsym,  esint}
\usepackage{array, mathtools, subfigure, esint, CJK}
\usepackage{makecell, multirow, ulem, bm}
\newcommand{\uat}[2]{\href{http://astrothesaurus.org/uat/#2}{#1 (#2)}}
          
\def\hmpc{\, {h {\rm Mpc}^{-1}}}           \def\mpch{\, {h^{-1} {\rm Mpc}}}

\def\dd{{\rm d}}                    
        
\def\vu{{\bf u}}          
\def\vx{{\bf x}}          
\shorttitle{Turbulence with wavelet for cosmic fluid}
\shortauthors{Wang \& He}
\begin{document}
\begin{CJK*}{UTF8}{gbsn}
\title{Turbulence revealed by wavelet transform: power spectrum and intermittency for the velocity field of the cosmic baryonic fluid}

\author[0000-0003-4064-417X]{Yun Wang (王云)}
\affiliation{College of Physics, Jilin University, Changchun 130012, China.}

\author[0000-0001-7767-6154]{Ping He (何平)}
\affiliation{College of Physics, Jilin University, Changchun 130012, China.}
\affiliation{Center for High Energy Physics, Peking University, Beijing 100871, China.}

\correspondingauthor{Yun Wang; Ping He}
\email{yunw@jlu.edu.cn; hep@jlu.edu.cn}

\begin{abstract}
We use continuous wavelet transform techniques to construct the global and environment-dependent wavelet statistics, such as energy spectrum and kurtosis, to study the fluctuation and intermittency of the turbulent motion in the cosmic fluid velocity field with the IllustrisTNG simulation data. We find that the peak scale of the energy spectrum define a characteristic scale, which can be regarded as the integral scale of turbulence, and the Nyquist wavenumber can be regarded as the dissipation scale. With these two characteristic scales, the energy spectrum can be divided into the energy-containing range, the inertial range and the dissipation range of turbulence. The wavelet kurtosis is an increasing function of the wavenumber $k$, first grows rapidly then slowly with $k$, indicating that the cosmic fluid becomes increasingly intermittent with $k$. In the energy-containing range, the energy spectrum increases significantly from $z = 2$ to $1$, but remains almost unchanged from $z = 1$ to $0$. We find that both the environment-dependent spectrum and kurtosis are similar to the global ones, and the magnitude of the spectrum is smallest in the lowest-density and largest in the highest-density environment, suggesting that the cosmic fluid is more turbulent in a high-density than in a low-density environment. In the inertial range, the energy spectrum's exponent is steeper than both the Kolmogorov and Burgers exponents, indicating more efficient energy transfer compared to Kolmogorov or Burgers turbulence.
\end{abstract}

\section{Introduction}
\label{sec:intro}

The turbulent motion of the cosmic baryonic fluid in large-scale structures of the universe has attracted increasing attention in cosmological studies over the last several decades. The physical origin of turbulence in cosmic baryonic fluids is complex and diverse, mainly including the following. Accretion through structure formation or hierarchical mergers should be able to generate and sustain turbulence in the intergalactic medium (IGM) and in galaxy clusters \citep{Subramanian2006, Bauer2012, Iapichino2017}, and the associated physical processes, such as the injection and amplification of vorticity by shock waves \citep[e.g.][]{Ryu2008, Porter2015, Vazza2017} or ram pressure stripping \citep[e.g.][]{Cassano2005, Subramanian2006, Roediger2007}, can also generate turbulence in clusters of galaxies. In addition, outflows or feedbacks from active galactic nucleus (AGN) jets, which inflate buoyant bubbles and eventually stir the intracluster medium (ICM), are expected to drive turbulence and heat cluster cores \citep{Bruggen2005, Iapichino2008b, Gaspari2011, Banerjee2014, Angelinelli2020}, and the supernova (SN)-driven galactic winds are also expected to generate turbulence in and around galaxies \citep[e.g.][]{Evoli2011, Iapichino2013}.

Observational evidence for turbulent motions in the ICM has been provided by direct detection of non-thermal broadening of the X-ray emission lines by the Hitomi satellite \citep{Hitomi2016}, and by indirect observations such as fluctuations in the magnetic field in the diffuse cluster radio sources \citep{Vogt2003, Murgia2004, Vogt2005, Ensslin2006, Bonafede2010, Vacca2010, Vacca2012}, fluctuations in the X-ray surface brightness or in pressure inferred from X-ray and Sunyaev-Zel'dovich effect maps \citep{Schuecker2004, Churazov2012, Gaspari2014, Zhuravleva2014, Walker2015, Khatri2016, Zhuravleva2018}, and the suppression of resonant line scattering in the X-ray spectra \citep{Churazov2004, Zhuravleva2013, Hitomi2018, Shi2019}. These effects can be used to detect and measure turbulence in the cosmic baryonic fluids. In addition to observations, the origin and evolution of turbulence in cosmic baryonic matter has been extensively studied using various cosmological hydrodynamical simulations \citep[e.g.][]{Norman1999, Dolag2005, Iapichino2008a, Lau2009, Zhu2010, Vazza2011, Gaspari2014, Miniati2014, Bruggen2015, Angelinelli2020}. 

There are a large number of problems relevant to the study of the formation and evolution of galaxies and galaxy clusters. Among these, the study of the heating mechanisms in the ICM and IGM is particularly important. If there is no sufficient heating mechanism, then there will be a phenomenon called cooling flow within the galaxy clusters, and there will also be an excess of star birth in the galaxy, which is known as the overcooling problem \citep{Voit2005}. In addition, a heating mechanism may also be responsible for the missing baryon problem \citep{Bregman2007}.

The main proposals to overcome this overcooling problem are the heating mechanisms such as SN or AGN feedback, and turbulence in the IGM or ICM. In the popular semi-analytical models of galaxy formation, the heating mechanism is mainly based on SN and AGN feedback, and turbulent heating is not considered \citep{Guo2011, Henriques2015}. However, some studies suggest that turbulent heating has a significant effect and should not be ignored. For example, \citet{Zhu2010} suggests that the turbulent pressure can be compared to the thermodynamic pressure of the baryonic gas. Furthermore, \citet{Zhuravleva2014} suggests that turbulent heating is sufficient to offset radiative cooling, and indeed appears to be locally balanced at each radius, which may be a key element in solving the gas cooling problem in cluster cores.

In a series of papers, we have carried out a number of studies on the turbulence of the cosmic baryon fluid, briefly summarised as follows. \citet{Hep2005} explores the dynamics and evolution of the baryonic gas in the universe, focusing on its velocity field and its interaction with the dark matter gravity, and discusses the implications of a Burgers fluid model for describing the dynamics of the IGM, highlighting the importance of considering heating and cooling processes in understanding the evolution of the IGM. Using the WIGEON hydrodynamical simulation \citep{Feng2004}, \citet{Hep2006} shows that the intermittency of the velocity field of the cosmic baryonic fluid at redshift $z = 0$ in the scale range from the Jeans length, roughly $0.1 - 0.3\mpch$, to about 16$\mpch$ can be described extremely well by She-Leveque’s universal scaling formula, and these results imply that the motion of the highly evolved cosmic baryonic fluid is similar to a fully developed turbulence. \citet{Yang2020} use updated WIGEON data \citep{Zhu2013} and study the turbulence-induced deviation of the spatial distributions between baryons and dark matter. They find that at $z = 0$, at the 1\% deviation level, the deviation scale is about 3.7$\mpch$ for the density field, while it is as large as 23$\mpch$ for the velocity field, a scale that falls into the weakly nonlinear regime of the density field for the structure formation paradigm. Their results also suggest that the effect of turbulence heating is comparable to that of these processes such as SN and AGN feedback. \citet{Yang2022} compare the results derived from IllustrisTNG and WIGEON simulations, and find that for the ratio of the density power spectrum between dark matter and baryonic matter, as scales become smaller and smaller, the power spectra for baryons are increasingly suppressed for WIGEON simulations, while for TNG simulations, the suppression stops at $k = 15 - 20 \hmpc$, and the power spectrum ratios increase when $k > 20 \hmpc$. These results indicate that turbulent effects can also have the consequence to suppress the power ratio between baryons and dark matter. These are of great importance for understanding the distribution and evolution of baryonic matter in the universe, as well as issues related to galaxy formation, and therefore turbulence in the cosmic baryonic fluid is worthy of in-depth study.

The wavelet analysis technique is a powerful tool, and the application of the wavelet transform technique to turbulence can be traced back more than three decades. In a review article, \citet{Farge1992} extensively discusses wavelet transforms as a powerful tool for achieving scale localization, allowing for the analysis of different scales of motion in turbulent flows. The article delves into the computation of local energy spectra, wavelet coefficients, and statistical properties, showing how wavelet analysis facilitates the investigation of coherent structures, intermittency, and other key aspects of turbulence. The paper details the application of wavelet transforms to turbulence analysis, emphasizing their ability to provide a localized and multi-scale view of the flow, leading to insights into coherent structures, intermittency, and statistical properties. \citet{Farge1992} emphasizes the advantages of wavelet analysis over traditional methods and provides concrete examples of its application to turbulent phenomena, providing a valuable resource for fluid mechanics researchers seeking a multi-scale perspective on turbulent flows.

The wavelet analysis method can also find its application in the study of cosmological turbulence. To name just a few that are relevant to our current interests. \citet{Schuecker2004} construct two-dimensional pressure maps from XMM-Newton observations of the Coma galaxy cluster using wavelet methods, which effectively suppress noise and reveal small-scale turbulence structures, helping to isolate and analyze pressure fluctuations, enabling the computation of their power spectra and providing evidence for turbulence in the ICM. \citet{Kowal2010} use wavelet analysis to decompose the velocity field of compressible magnetohydrodynamic turbulence into Alv$\acute{\rm e}$n, slow, and fast modes, and investigate in detail the turbulence properties such as spectra, anisotropy, scaling exponents and intermittency, showing that these properties are influenced by both the sonic and Alv$\acute{\rm e}$n Mach numbers. \citet{Shi2018} use a novel wavelet analysis method to study the radial dependence of the ICM turbulence spectrum, and find that faster turbulence dissipation in the inner high-density regions causes the turbulence amplitude to increase with radius. They also find that the ICM turbulence at all radii decays in two phases after a major merger, i.e. an early fast-decay phase and a slow secular-decay phase.

In a series of works, we have used the continuous wavelet transform (CWT) to construct the wavelet statistics such as the wavelet power spectrum (WPS), wavelet cross-correlation and wavelet bicoherence, as well as the environment-dependent WPS (env-WPS), and analyzed the clustering and non-Gaussian properties of large-scale structures in the universe using the cosmological simulation data \citep{Wang2021, Wang2022a, Wang2022b}. In a recent paper \citep{Wang2024}, we mainly use the env-WPS to study how baryonic effects vary with scale and local density environment. In addition, we also complete a comparative study of several fast algorithms for one-dimensional (1D) CWT \citep{Wang2023}.

In this work, we apply the CWT techniques to the IllustrisTNG simulation data to study the turbulent motion of the cosmic baryonic fluid in the large-scale structures of the universe. The paper is organized as follows. In Section~\ref{sec:meth}, we briefly introduce our methods and the simulation data used in this work. In Section~\ref{sec:result}, we present the results of our work. In Section~\ref{sec:concl}, we present the summary and conclusions of the paper.

\section{Methods and Data}
\label{sec:meth}

\subsection{Continuous Wavelet Transform and Power Spectrum}
\label{sec:wvt}

For a random field with the zero mean value, say the 3-dimensional (3D) velocity field of the cosmic baryonic fluid $\vu$, its isotropic CWT $\tilde{\vu}(w,\vx)$ is obtained by convolution with the wavelet function $\Psi$ as 
\begin{flalign}
\label{eq:wvt}
\tilde{\vu}(w,\vx) = \int\vu(\boldsymbol{\tau})\Psi(w, \vx - \boldsymbol{\tau})\dd^3\boldsymbol{\tau}.
\end{flalign}
Throughout this work, we use the so-called 3D isotropic cosine-weighted Gaussian-derived wavelet (CW-GDW), which can achieve good localization in both spatial and frequency space simultaneously \citep{Wang2022b, Wang2024}. It is defined in real space as
\begin{flalign}
\label{eq:CW-GDW_3D}
\Psi(w,\mathbf{x}) &= w^\frac{3}{2}\Psi(wr) \nonumber \\
& = C_\mathrm{N}w^\frac{3}{2}\Big[ (4-w^2r^2)\cos wr \nonumber \\ 
& \quad + 2\big(\frac{1}{wr}-wr\big)\sin wr \Big]e^{-\frac{1}{2}w^2r^2},
\end{flalign}
with its Fourier transform
\begin{flalign}
\label{eq:CW-GDW_3D_k}
\hat\Psi(w,\mathbf{k}) & = w^{-\frac{3}{2}}\hat\Psi(k/w) \nonumber \\
& =(2\pi)^\frac{3}{2}C_\mathrm{N}w^{-\frac{3}{2}}\Big(\frac{k}{w}\Big)\Big[\Big(\frac{k}{w}\Big)\cosh\frac{k}{w} \nonumber \\
& \quad-\sinh\frac{k}{w}\Big]e^{-\frac{1}{2}(1+k^2/w^2)},
\end{flalign}
where $r=|\mathbf{x}|$, $k=|\mathbf{k}|$, and $C_\mathrm{N}=\frac{2}{\pi^{3/4}}\sqrt{2e/(9+55e)}$ is the normalization constant such that $\int |\Psi(\mathbf{x})|^2\mathrm{d}^3\mathbf{x}=1$.

In many cases, we compare the results of the CWT with those of the Fourier transform, such as comparing the global WPS with the Fourier power spectrum (FPS). However, the wavelet scale $w$ and the Fourier wavenumber $k$ are not simply equal, but have a correspondence $w = c_w k$, where $c_w \simeq 0.3883$ for the 3D isotropic CW-GDW (see Appendix~\ref{sec:fps_vs_wps}).

\begin{figure*}
\centerline{\includegraphics[width=0.85\textwidth]{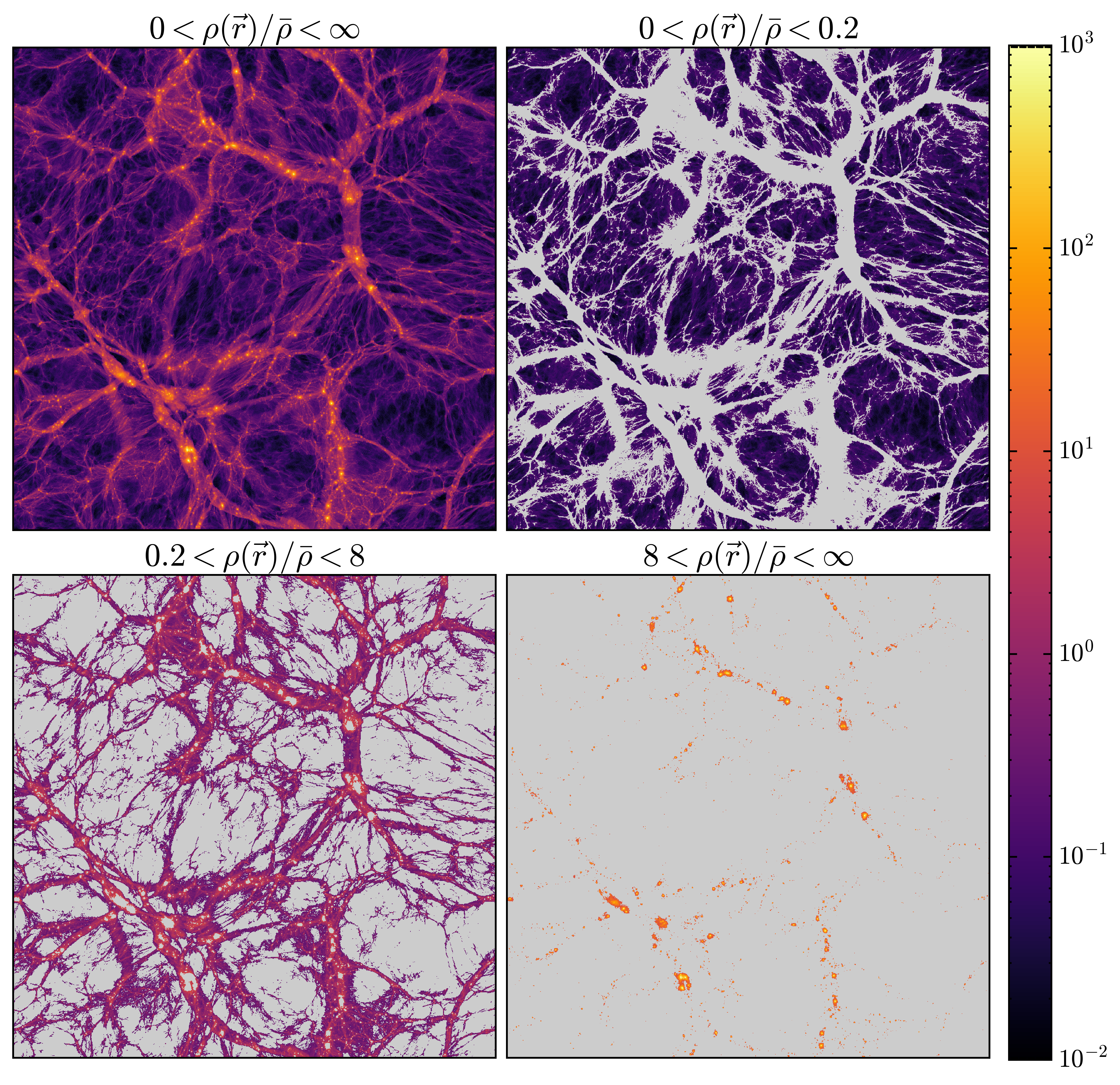}}
\caption{A two-dimensional slice of the IllustrisTNG100-1 dark matter density field at $z=0$ (top left), and spatial regions corresponding to the density ranges labelled in these panels. The slice covers a $75\times75 h^{-2} {\rm Mpc}^2$ area with a thickness of $0.3\mpch$. The grey background indicates the region outside the specified density range. Fourier analysis can only be applied to the regular region, such as a square or a cube (top left panel), whereas wavelet analysis can be applied to regions of any shape (all four panels).}
\label{fig:density_fields}
\end{figure*}

With the CWT $\tilde{\vu}(w, \vx)$, we can define the env-WPS of the velocity field as
\begin{flalign}
\label{eq:env_wps}
\tilde{P}_\vu(w,\Delta) \equiv \Big\langle |\tilde{\vu}(w,\vx)|^2 \Big\rangle_{\vx \in V_\Delta} = \frac{1}{N_V} \sum_{i=1}^{N_V} |\tilde{\vu}(w,\vx_i)|^2, 
\end{flalign}
where the environment is specified with the dark matter density contrast $\Delta=\rho_{\rm dm}/\bar{\rho}_{\rm dm}$. For the environment $\Delta$, we specify a density interval $\delta_1 < \Delta < \delta_2$ as the density-limiting condition, and determine all the spatial regions $V_1$, $V_2$, $V_3$, ..., with $V_\Delta = V_1 \cup V_2 \cup V_3 ...$, such that the density-limiting condition is satisfied within $V_\Delta$. The spatial coordinate $\vx$ runs over all the spatial region $V_\Delta$, and we count the number of all the coordinates as $N_V$. In this way, we obtain the env-WPS of Equation~(\ref{eq:env_wps}). If the density interval is taken as $0 < \Delta < \infty$, then we obtain the global WPS as,
\begin{flalign}
\label{eq:global_wps}
\tilde{P}_\vu(w) \equiv \Big\langle |\tilde{\vu}(w,\vx)|^2 \Big\rangle_{{\rm all}\ \vx},
\end{flalign} 
which is actually calculated by averaging over all spatial positions. It can be seen that the global WPS $\tilde{P}_\vu(w)$ is related to the FPS $P_\vu(k)$ as \citep{Wang2022a}
\begin{flalign}
\label{eq:gwps_fps}
\tilde{P}_\vu(w) = \frac{1}{2\pi^2}\int^{+\infty}_{0}P_\vu(k) \hat\Psi(w, k)^2 k^2 \dd k.
\end{flalign}

One can see that the global WPS is the wavelet-weighted average of the FPS in $k$ space. The former therefore has a similar but smoother shape to the latter. Obviously, the wider the bandwidth of the wavelet function, the smoother the global WPS. If the bandwidth is narrower, the global WPS is closer to the FPS.

Integrating both sides of Equation~(\ref{eq:gwps_fps}) with respect to $w$, we can obtain an approximate relationship with the correspondence $w = c_w k$ as
\begin{flalign}
\label{eq:wfps}
\tilde{P}_\vu(k) \approx \frac{I_\Psi}{2\pi^2 c_w} P_\vu(k),
\end{flalign}
where $I_\Psi=\int^{+\infty}_0 k |\hat\Psi(k)|^2 \dd k$ is a constant depending on the wavelet function. Therefore, the global WPS and the FPS differ by only a scaling factor that depends on the specific wavelet function selected. Obviously, for the power-law power spectrum $P_\vu(k) \propto k^n$, $\tilde P_\mathbf{u}(k)$ preserves the same power index. For the 3D isotropic CW-GDW used in this work, $I_\Psi/(2\pi^2 c_w) \approx 1.0754$, at which point the shape and size of the global WPS and FPS are very close (see Appendix~\ref{sec:fps_vs_wps}). However, due to the simultaneous spatial and frequency localization of the CWT, we can specify any shape of the region for integration, such as a density-bounded region. Therefore, we can define the env-WPS as in Equation~(\ref{eq:env_wps}). Because the env-WPS can be calculated in spaces of arbitrary shapes, whereas the FPS is only restricted to regular shapes like squares or cubes, the techniques of wavelet analysis are superior to those of Fourier analysis, as demonstrated in Figure~\ref{fig:density_fields}.

Incidentally, it is not necessary to use anisotropic wavelets to compute the velocity WPS. In \citet{Wang2022b}, we compute the matter WPS using both 3D isotropic and anisotropic wavelet, and find that they yield almost identical results. For details of the CWT techniques that we have developed, please refer to \citet{Wang2021}, \citet{Wang2022a}, \cite{Wang2022b, Wang2023} and \citep{Wang2024}.

\begin{figure*}
\centerline{\includegraphics[width=1.0\textwidth]{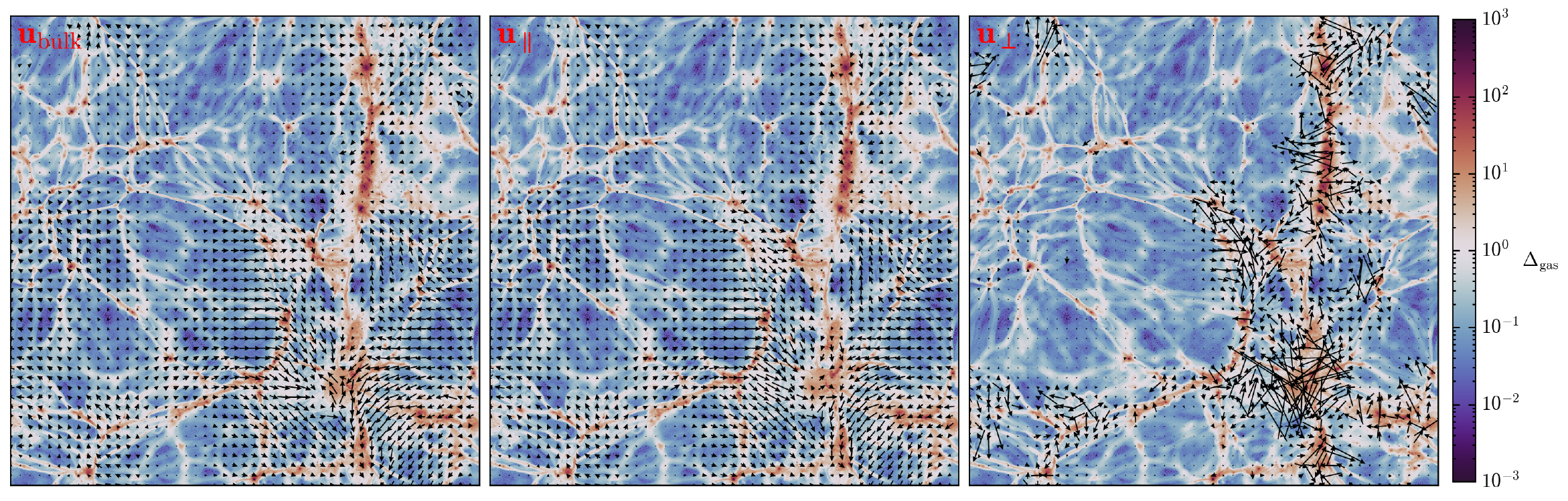}}
\caption{A two-dimensional slice of the velocity field for the TNG50-1 data at $z=0$.  The slice covers a $35\times 35 h^{-2} {\rm Mpc}^2$ area with a thickness of $0.18\mpch$. The line of sight is along the ${\rm z}$-axis, and only the velocity in the ${\rm x-y}$ direction is displayed. The panels on the left, middle, and right correspond to the bulk velocity $\vu_{\rm bulk}$, the compressive component $\vu_\parallel$ and the solenoidal component $\vu_\bot$, respectively. The background represents the density contrast $\Delta_{\rm gas}\equiv\rho_{\rm gas}(\vx)/\bar{\rho}_{\rm gas}$ of the baryonic fluid.}
\label{fig:velocity_map}
\end{figure*}

\subsection{Turbulence in the Cosmic Baryonic Fluid}
\label{sec:turb}

As mentioned in Section~\ref{sec:intro}, the cosmic baryonic fluid is characterized by turbulent motion. In this work, we investigate in which regions, to what extent, and on which scales the cosmic baryonic fluid is turbulent.

\subsubsection{Filtering Out the Bulk Flow}
\label{sec:debulk}

Although it is difficult to define turbulence \citep{Davidson2015}, we can still provide a description of turbulent flow in fluids and distinguish between turbulent and bulk flow.

In general, a velocity field of the fluid $\vu$ is a superposition of the turbulent velocity and the bulk velocity, i.e. $\vu = \vu_{\rm turb} + \vu_{\rm bulk}$. Bulk flow is a flow state in which the fluid particles move in a relatively orderly manner, and the flow velocity and direction are relatively uniform in a given cross-section. In contrast, turbulent flow refers to a chaotic, disordered flow state in which the fluid particles move irregularly and randomly, and usually occurs when the fluid velocity is high, the pressure gradient is large, or the Reynolds number exceeds a certain critical value. Turbulent flow has the characteristics of rapid mixing, strong diffusion, and good heat transfer. Therefore, it is necessary to develop an approach to separate the turbulent flow from the bulk motion \citep[e.g.][]{Dolag2005, Vazza2012, ZuHone2013, Shi2018, Angelinelli2020, Valles-Perez2021a}. 

In this study, we utilize the code based on the iterative multi-scale filtering approach of \citet[][see the Appendix of their paper]{Vazza2012} to extract turbulent motions from the velocity field of the cosmic fluid. Specifically, we apply no weighting scheme when calculating the local mean velocity field, i.e. $w_i=1$ in Equation~(3) of \citet{Vazza2012}. To use the code, we set the relevant parameters as \texttt{eps=0.05}, \texttt{epssk=0.5} and \texttt{nk=16}.

\subsubsection{Intermittency of Turbulence}
\label{sec:intermittent}

In general, there are two types of turbulence in fluids, Kolmogorov turbulence and Burgers turbulence. Kolmogorov turbulence is essentially excited by subsonic motion of the flow, and characterized by eddies of different scales \citep{Bauer2012}. In the fully developed state, the fluid satisfies $\nabla\cdot\vu=0$, and its kinetic energy spectrum follows the Kolmogorov's $k^{-5/3}$ scaling-law. Burgers turbulence is essentially excited by supersonic motion of the flow \citep{Boldyrev2004, Federrath2010, Bauer2012} and characterized by shock waves \citep{Konstandin2015}, which gives $k^{-2}$ for the kinetic energy spectrum at steady state when ${\rm Re}$, the Reynolds number, satisfies ${\rm Re} \gg k L_{\rm c} \gg 1$, where $L_{\rm c}$ is some correlation length \citep{Balkovsky1997}. The turbulence in the cosmic baryonic fluid is composed of both Kolmogorov turbulence and Burgers turbulence, but it is not simply a mixture of the two types of turbulence, as will be discussed later.

In turbulence, the presence of shocks leads to a strong intermittency, and therefore the turbulent energy dissipation in space is also intermittent. Intermittency in turbulence refers to the phenomenon where fluctuations in velocity, pressure or other flow properties occur in a sporadic or irregular manner. Intermittency arises because these fluctuations are not uniformly distributed in space and time. Studying intermittency helps in understanding how energy cascades from large to small scales and dissipates. Usually, the deviation of the tails of the probability density function (PDF) from Gaussian is seen as a manifestation of intermittency, with a highly unusual scaling for the moments of the velocity differences \citep{Balkovsky1997}. Therefore, to characterize the intermittency of turbulence in the cosmic fluid, as in \citet{Meneveau1991, Angulo2023}, we define the wavelet global kurtosis (or flatness) of the cosmic velocity field as
\begin{flalign}
\label{eq:gkuw}
{\rm {\tilde K}}_\vu(k) \equiv \frac{\left<\Delta \tilde{u}(k,\vx)^4\right>_{{\rm all}\ \vx}}{\left<\Delta \tilde{u}(k,\vx)^2\right>^2_{{\rm all}\ \vx}},
\end{flalign}
and the wavelet environment-dependent (env-) kurtosis as
\begin{flalign}
\label{eq:lkuw}
{\rm \tilde{K}}_\vu(k, \delta) \equiv \frac{\left<\Delta \tilde{u}(k,\vx)^4\right>_{\delta(\vx)=\delta}}{\left<\Delta \tilde{u}(k,\vx)^2\right>^2_{\delta(\vx)=\delta}},
\end{flalign}
in which $\tilde{u}(k,\vx)$ is the CWT of the velocity component along a given axis, and $\Delta \tilde{u}(k,\vx) = \tilde{u}(k,\vx) - \left< \tilde{u}(k,\vx) \right>$. For a scalar Gaussian field, the ${\rm kurtosis}=3$, whereas for the turbulent velocity field of fluids, its PDF is generally a long-tailed distribution, with ${\rm kurtosis}>3$. 

Kurtosis can be thought of as a measure of the intermittency of the turbulence. The greater the kurtosis, the more intermittent the cosmic flow should be.

Note that \citet{Farge1992} use the local wavelet energy spectrum of the velocity field to define the local intermittency of turbulent flow, see their Equation-(59). We do not use this definition as we have already used env-WPS to perform the spectral analyses throughout this work.

\subsubsection{Ratio of Power Spectra of the Two Modes}

The velocity field of the cosmic baryonic fluid $\vu$, can be separated by the Helmholtz-Hodge decomposition \citep{Arfken2005} into a compressive (or longitudinal) part `${\parallel}$' and a solenoidal (or transverse) part `$\bot$', as $\vu = \vu_\parallel + \vu_\bot$, where the compressive part $\vu_\parallel$ satisfies $\nabla \times \vu_\parallel = 0$, and the solenoidal part $\vu_\bot$ satisfies $\nabla \cdot \vu_\bot = 0$, respectively. In Figure~\ref{fig:velocity_map}, we present a visualization of the velocity field for the TNG50 data at $z=0$.

For the divergence $d \equiv \nabla\cdot\vu$ and the vorticity $\boldsymbol{\omega} \equiv \nabla\times\vu$ of the velocity field of $\vu$, it is easy to show that 
\begin{flalign}
\label{eq:pdpw}
P_d(k) = k^2 P_\parallel(k),  \qquad P_\omega(k) = k^2P_\bot(k),
\end{flalign}
in which $P_d(k)$, $P_\omega(k)$, $P_\parallel(k)$ and $P_\bot(k)$ are the FPS of $d$, $\boldsymbol{\omega}$, the velocity component $\vu_\parallel$ and $\vu_\bot$, respectively. 

The small-scale compression ratio can be defined as follows \citep{Kida1990, Schmidt2009} 
\begin{flalign}
\label{eq:rcs}
r_{\rm CS} \equiv \frac{\left< d^2 \right>}{\left< d^2 \right> + \left< \omega^2 \right>}.
\end{flalign}
This ratio quantifies the relative importance of the compressive and solenoidal modes in a flow \citep{Iapichino2011}. Inspired by this compression ratio, we define the FPS ratio of the solenoidal mode to the total turbulence as
\begin{flalign}
\label{eq:fps_ratio}
Q(k) \equiv \frac{P_\omega(k)}{P_d(k) + P_\omega(k)} = \frac{P_\bot(k)}{P_\parallel(k) + P_\bot(k)}.
\end{flalign}
It can be seen that $Q(k)$ is superior to $r_{\rm CS}$ as it is scale dependent. Similar to the FPS ratio in Equation~(\ref{eq:fps_ratio}), we can also define the global WPS ratio as
\begin{flalign}
\label{eq:gwps_ratio}
\tilde{Q}(k) \equiv \frac{\tilde{P}_\omega(k)}{\tilde{P}_d(k) + \tilde{P}_\omega(k)} = \frac{\tilde{P}_\bot(k)}{\tilde{P}_\parallel(k) + \tilde{P}_\bot(k)},
\end{flalign}
in which $\tilde{P}_d(k)$, $\tilde{P}_\omega(k)$, $\tilde{P}_\parallel(k)$, and $\tilde{P}_\bot(k)$ are global WPS of $d$, $\boldsymbol{\omega}$, the velocity component $\vu_\parallel$ and $\vu_\bot$, respectively, computed in the same way as Equation~(\ref{eq:global_wps}), and the correspondence $w = c_w k$ is used to convert $w$ to $k$.

As mentioned above, the turbulence in the cosmic fluid consists not only of Kolmogorov turbulence, but also of Burgers turbulence. Therefore, it is not proper to treat the turbulent energy spectrum as consisting only of the solenoidal component \citep[cf.][]{Ryu2008}, and it is better to take $P_d(k) + P_\omega(k)$, or its wavelet counterpart $\tilde{P}_d(k) + \tilde{P}_\omega(k)$, as the total energy spectrum. Thus $Q(k)$ and $\tilde{Q}(k)$ reflect the solenoidal fraction of the total turbulence energy as a function of $k$.

With the env-WPS of the velocity field in Equation~(\ref{eq:env_wps}), we can also define the env-WPS ratio as
\begin{flalign}
\label{eq:lwps_ratio}
\tilde{Q}(k, \delta) \equiv \frac{\tilde{P}_\omega(k, \delta)}{\tilde{P}_d(k, \delta) + \tilde{P}_\omega(k, \delta)} = \frac{\tilde{P}_\bot(k, \delta)}{\tilde{P}_\parallel(k, \delta) + \tilde{P}_\bot(k, \delta)},
\end{flalign}
in which $\tilde{P}_d(k, \delta)$, $\tilde{P}_\omega(k, \delta)$, $\tilde{P}_\parallel(k, \delta)$ and $\tilde{P}_\bot(k, \delta)$ are the env-WPS of $d$, $\boldsymbol{\omega}$, the velocity component $\vu_\parallel$ and $\vu_\bot$, respectively. $\tilde{Q}(k, \delta)$ is the ratio of the solenoidal component to the total energy as a function of $k$ and the environment $\delta$.

\begin{figure*}
\centerline{\includegraphics[width=0.975\textwidth]{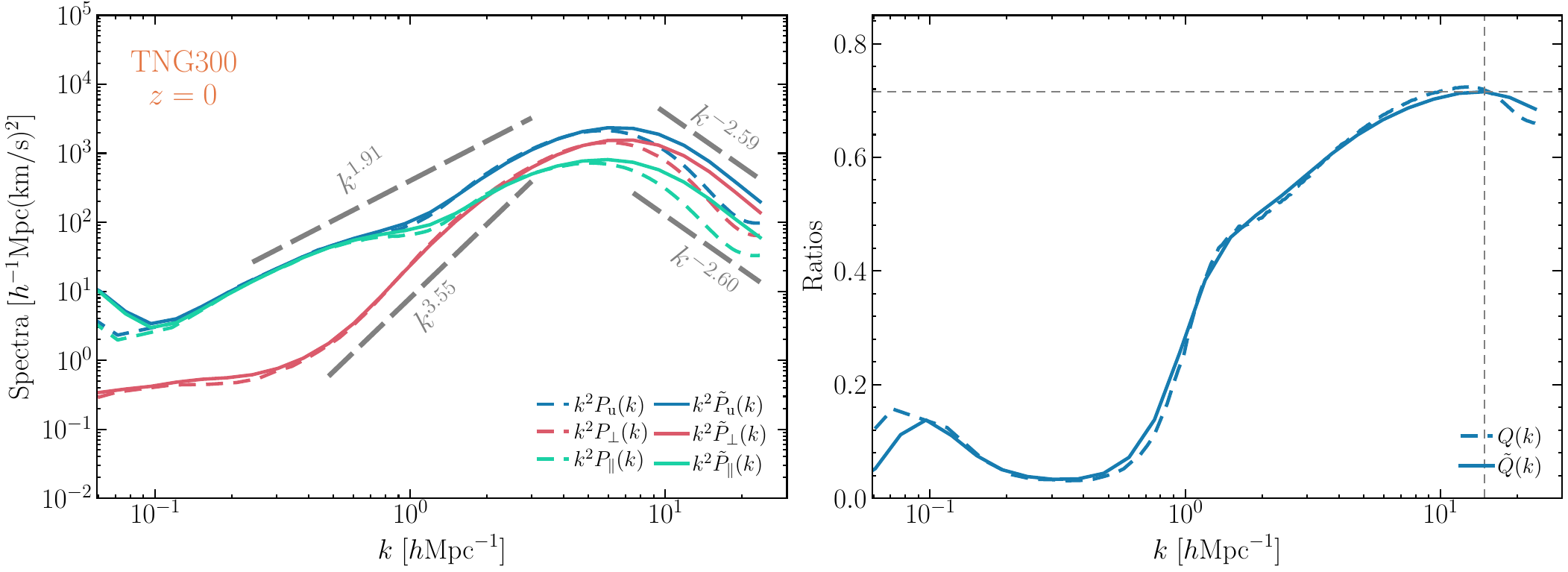}}
\centerline{\includegraphics[width=0.975\textwidth]{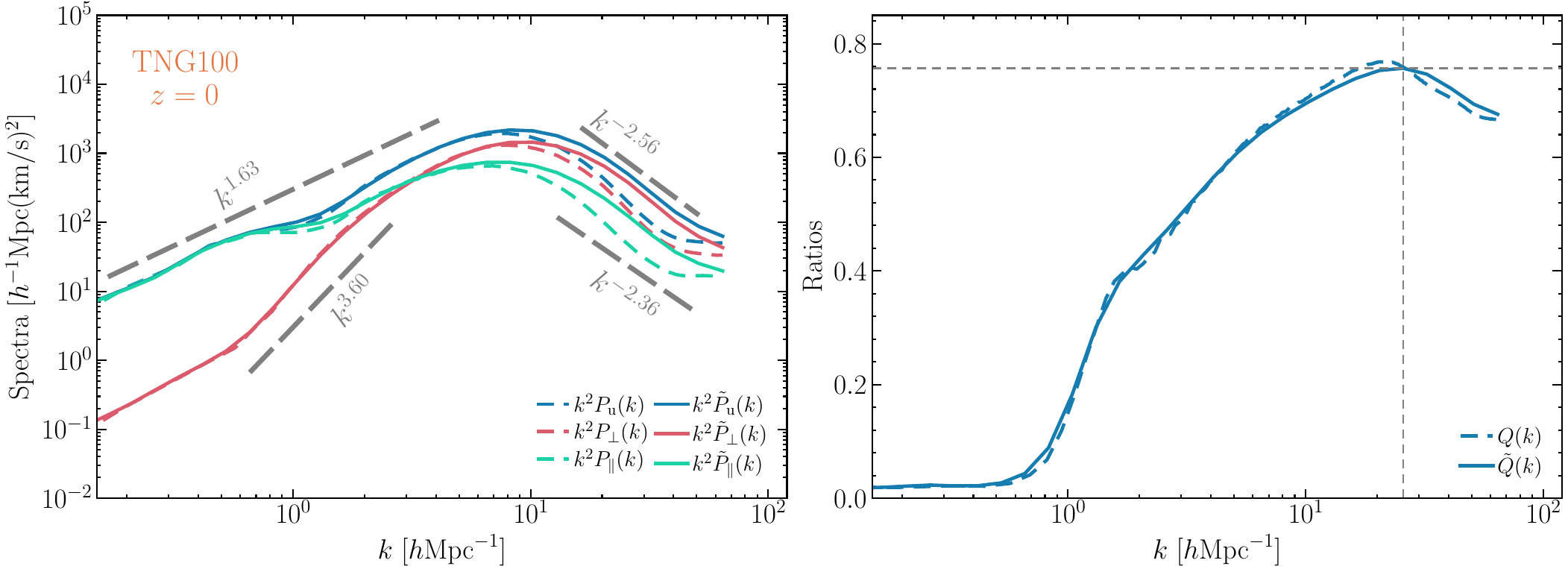}}
\centerline{\includegraphics[width=0.975\textwidth]{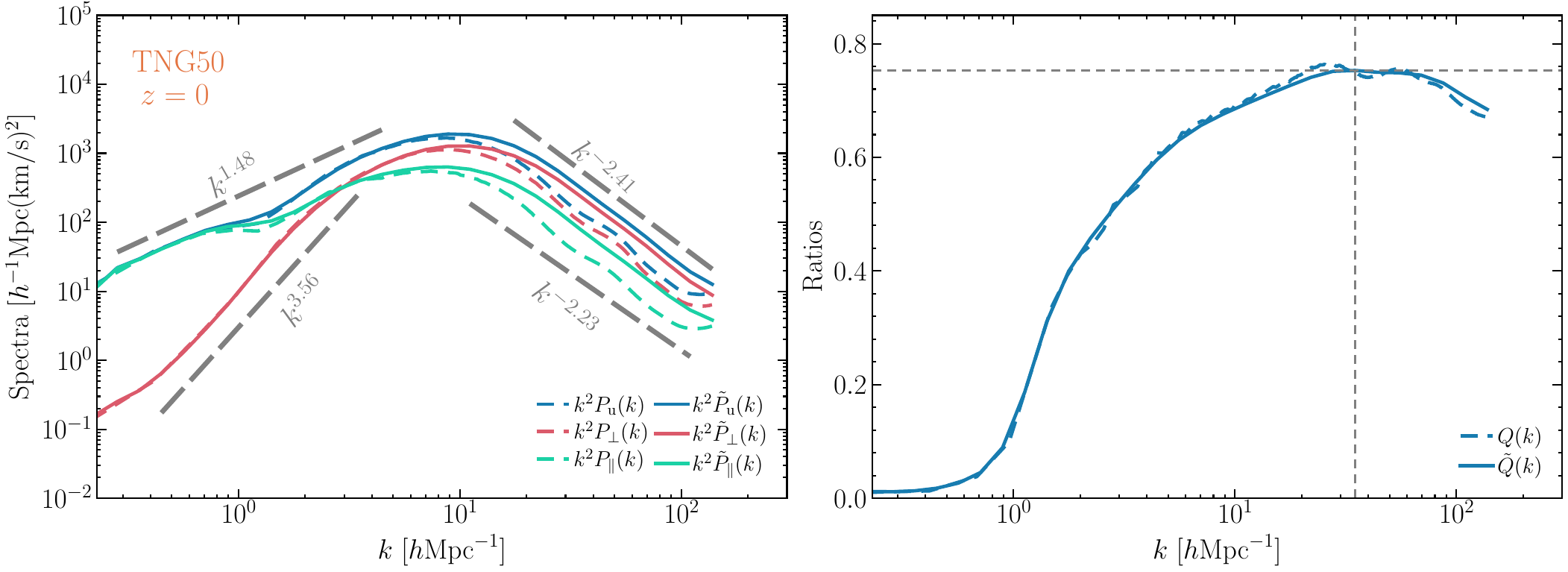}}
\caption{Both Fourier and global wavelet energy spectra at $z=0$ for the velocity field $\vu$, its `$\parallel$' and `$\bot$' component of the cosmic fluid (left column), with the corresponding spectral ratios (right column). From top to bottom are the TNG300, TNG100, and TNG50 data, respectively. To aid the eye, the relevant power-laws are indicated in the figure.}
\label{fig:TNG_velocity_globalwps}
\end{figure*}

\subsection{Simulation Data}
\label{sec:data}

We utilize data from the IllustrisTNG simulations (TNG for short) \citep{Pillepich2018a, Springel2018, Marinacci2018, Nelson2018, Naiman2018, Nelson2019}, focusing on the samples TNG300-1, TNG100-1, and TNG50-1. These samples correspond to simulated universes of sizes $205\mpch$, $75\mpch$ and $35\mpch$, respectively. TNG simulations are comprehensive, encompassing large-scale gravo-magnetohydrodynamical processes within the cosmos, and are executed using the moving-mesh code AREPO \citep{Springel2010}. This code employs a second-order accurate Godunov-type scheme for the hydrodynamic equations on a dynamically unstructured mesh, improving the fidelity of both supersonic and subsonic fluid turbulence simulations. \citet{Bauer2012} demonstrated AREPO's superiority in accurately replicating turbulence, achieving Kolmogorov-like scaling laws for density, velocity, and vorticity power spectra, aligning with theoretical expectations from the fully developed isotropic turbulence.

In addition to gravitational and hydrodynamic calculations, the TNG simulations incorporate a comprehensive set of physical processes \citep{Pillepich2018b, Nelson2019}, which include: (1) the formation and evolution of stars, (2) the associated metal enrichment and loss of mass, (3) cooling processes for primordial and metal-enriched gases, (4) the impact of SN on the pressurization of the interstellar medium, without resolving individual SN events, (5) feedback from stars, manifesting as galactic winds powered either by SN or stars in the asymptotic giant branch (AGB), (6) the formation and growth of supermassive black holes, accompanied by energy feedback from AGN in both high-accretion quasar and low-accretion kinetic wind phases, and (7) the role of magnetic fields in cosmic structures.

Given the abundance of the physical processes included in the TNG simulations and AREPO's excellent performance in hydrodynamical computations, the TNG data are well suited for studying cosmic fluid turbulence.

\section{Results}
\label{sec:result}

We use the piecewise cubic spline (PCS) scheme \citep{Sefusatti2016} to assign all the simulation particles into a $1536^3$ grid, and with this grid, we can use FFT to compute the FPS and WPS \citep[see][for the numerical pipeline]{Wang2024}. 

\begin{figure*}[htb]
\centerline{\includegraphics[width=0.975\textwidth]{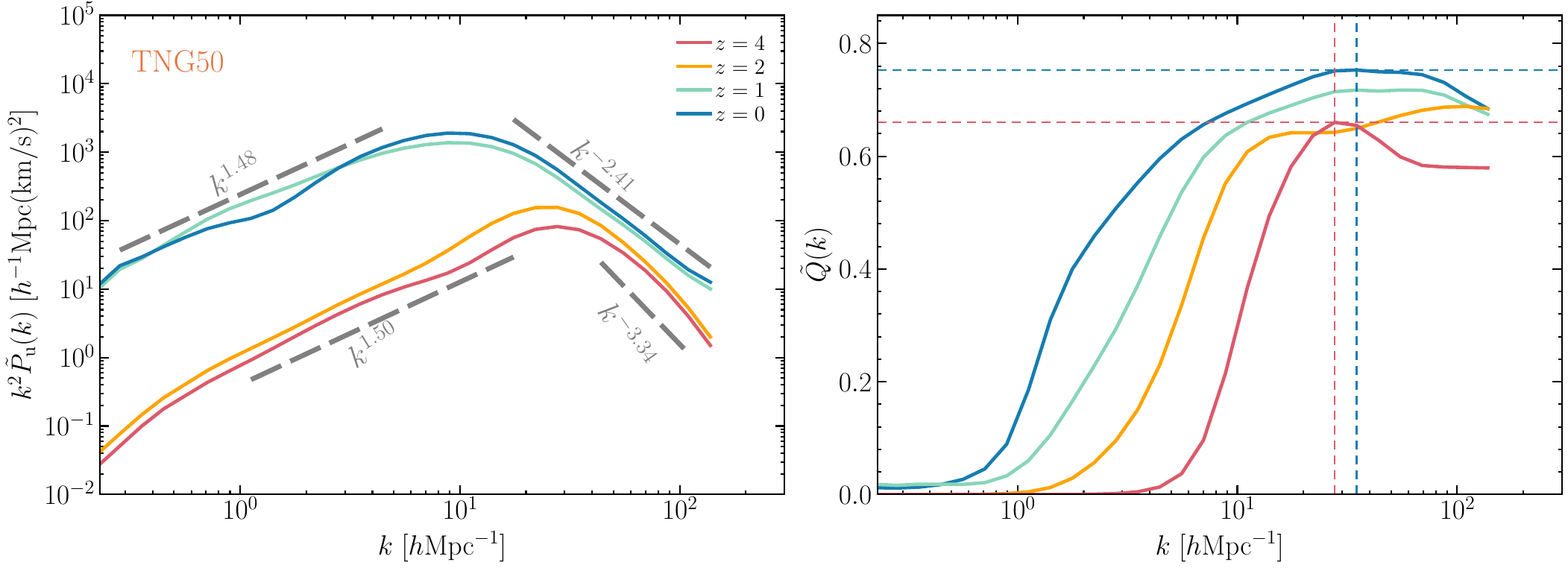}}
\caption{The $z$-evolution of the energy spectrum  (left panel) and the spectral ratio (right panel) for the TNG50 simulation. Similar to Figure~\ref{fig:TNG_velocity_globalwps}, the relevant power-laws are indicated in the figure.}
\label{fig:TNG50_velocity_globalwps_z0_4}
\end{figure*}

\subsection{The Global Results}
\label{sec:global_results}

As mentioned in Section~\ref{sec:debulk}, in all the following computations, the bulk flow has been removed from the velocity field using the approach of \citet{Vazza2012}.

In Figure~\ref{fig:TNG_velocity_globalwps}, we show both the FPS and the global WPS at $z=0$ for the velocity field $\vu$ and its `$\parallel$' and `$\bot$' components of the cosmic fluid, with the corresponding spectral ratios $Q(k)$ and $\tilde{Q}(k)$. From top to bottom are the results for the TNG300, TNG100, and TNG50 simulations, respectively. As analyzed in Section~\ref{sec:wvt}, the global WPS is almost the same as the FPS, just with a slightly higher factor. The ratios $Q(k)$ and $\tilde{Q}(k)$ also give almost the same results. Indeed, we observe that the Fourier analysis and the global wavelet analysis give consistent results, suggesting that the wavelet analysis is valid.

It can be seen that there are peaks in all energy spectra of the velocity $\vu$ of the cosmic fluid, and the peak scales are $k_{\rm S-peak} = 6.0,\ 8.2$ and $8.8\hmpc$, i.e. the length scale $1.0$, $0.76$ and $0.71\mpch$ for TNG300, TNG100 and TNG50, respectively. It seems that the peak scales are roughly $1/5 - 1/3$ of the sizes of virialized cluster halos. At large scales, i.e. $k < k_{\rm S-peak}$, the energy spectra are increasing functions of $k$, and the turbulent velocity $\vu_{\rm turb}$ is mostly dominated by its compressive component. Within this scale range, the gravitational collapse converts the potential energy of the structure into kinetic energy of the bulk flow, and the bulk flow further decays into turbulent flows - first in the compressive mode and, with increasing $k$, mostly in the solenoidal mode, as can also be seen from the spectral ratios $Q(k)$ and $\tilde{Q}(k)$. The transition scales of the two modes can be determined by setting $\tilde{Q}(k_{\rm trans})$ or $Q(k_{\rm trans}) = 1/2$, and thus we get $k_{\rm trans} = 1.9,\ 2.9$ and $2.7\hmpc$ for TNG300, TNG100 and TNG50 respectively.

\begin{figure}
\centerline{\includegraphics[width=0.5\textwidth]{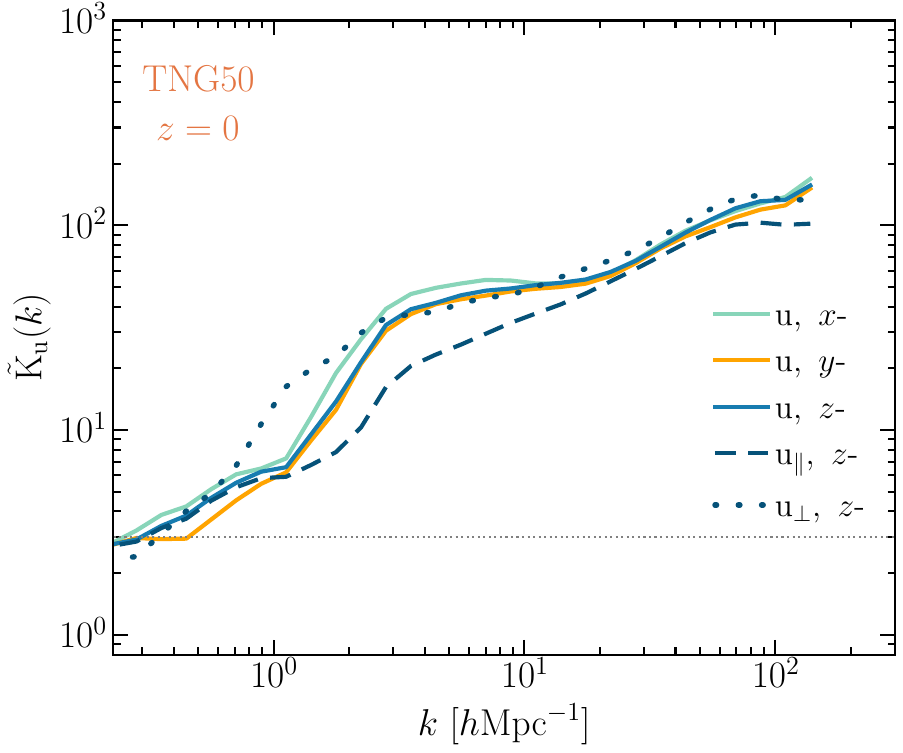}}
\caption{The global kurtosis of the velocity field $\vu$, $\vu_\parallel$ and $\vu_\bot$ of the TNG50 data at $z=0$, respectively. The curves show the $x$-, $y$-, $z$-direction of $\vu$ and the $z$-direction of $\vu_\parallel$ and $\vu_\bot$ respectively. The horizontal dotted line shows the ${\rm kurtosis} = 3$.}
\label{fig:TNG50_globalkurt_z0_xyz}
\end{figure}

According to the analysis by \citet{Zhu2010}, the baroclinic term in their dynamical equation~(1) drives the growth of vorticity over time, which becomes more effective during the non-linear evolutionary stage. We observe that the WPS of the solenoidal mode increases significantly faster with $k$ compared to that of the compressive mode if $k < k_{\rm S-peak}$. This is likely the reason for the dips observed at $k\sim1\hmpc$. 

One can see that the energy spectra drop rapidly at $k > k_{\rm S-peak}$. As shown in the figure, the exponent of the energy spectrum within this scale range is $-2.59$, $-2.56$ and $-2.41$ for TNG300, TNG100 and TNG50 respectively, which are all steeper than not only the Kolmogorov exponent $-5/3$ but also the Burgers turbulence exponent $-2$. As discussed in Section~\ref{sec:intermittent}, Kolmogorov turbulence is excited by subsonic motion of the flow, characterized by eddies of different scales, and hence the $\vu_\bot$ component is mostly responsible for Kolmogorov turbulence, while Burgers turbulence is excited by supersonic motion, characterized by shock waves, and hence the $\vu_\parallel$ component is mostly responsible for Burgers turbulence. The mechanism of energy dissipation is different for Kolmogorov and Burgers turbulence. In Kolmogorov turbulence, kinetic energy is cascaded from larger vortices to smaller and smaller vortices and then dissipated into thermal energy below the dissipation scale, whereas in Burgers turbulence, kinetic energy can be not only passed from large-scale to small-scale eddies, but also directly dissipated into thermal energy by shock waves at any scale in the inertial range. Hence, we conclude that a steeper energy spectrum results in a more efficient energy transfer in Burgers turbulence compared to Kolmogorov turbulence, attributed to the additional dissipation from kinetic to thermal energy in the inertial range. This conclusion can be extended to the general case: turbulence with a steeper energy spectrum generally exhibits a more efficient energy transfer than turbulence with a less steep spectrum. See Appendix~\ref{sec:relation_exp_etr} for a detailed discussion about the relationship between the energy spectrum's exponent and the rate of energy transfer.

We notice that there are also peaks in the spectral ratios $Q(k)$ and $\tilde{Q}(k)$. The peak scales $k_{\rm Q-peak} = 14.9,\ 25.7$ and $34.9\hmpc$ for TNG300, TNG100 and TNG50 respectively. The power of the $\vu_\bot$ component drops rapidly beyond $k_{\rm Q-peak}$.

\begin{figure}
\centerline{\includegraphics[width=0.5\textwidth]{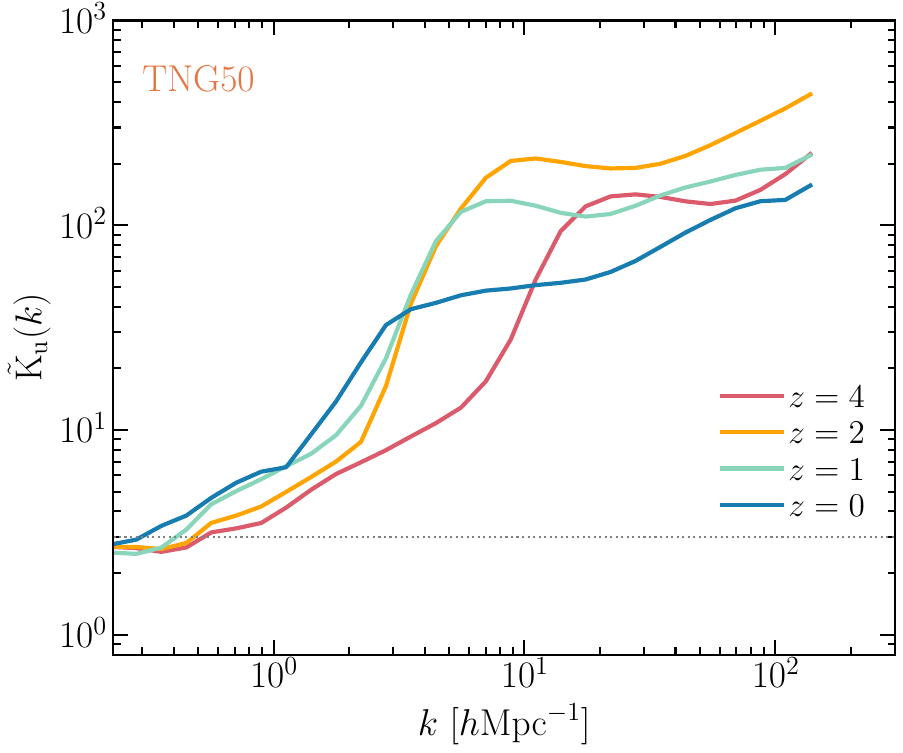}}
\caption{$z$-evolution of the global kurtosis for the TNG50 data. The horizontal dotted line shows the ${\rm kurtosis} = 3$.}
\label{fig:TNG50_globalkurt_z0_4_z}
\end{figure}

\begin{figure*}
\centerline{\includegraphics[width=0.975\textwidth]{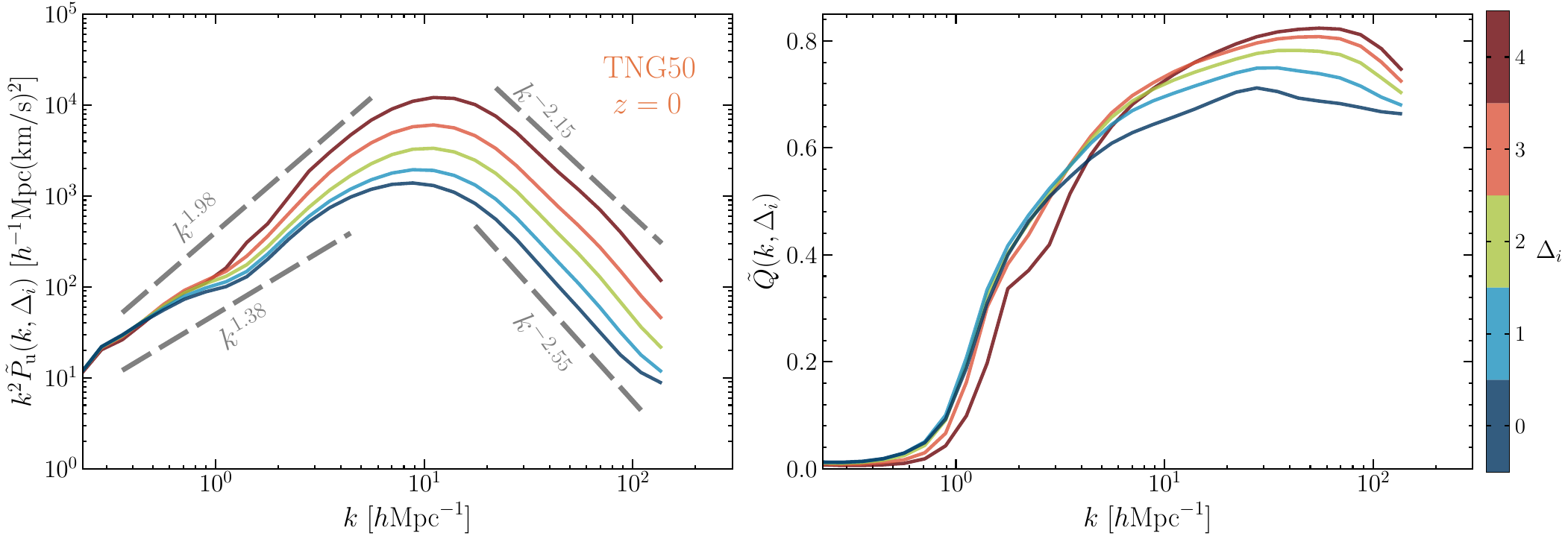}}
\caption{The environment-dependent wavelet energy spectra of the TNG50 simulation at $z=0$ for the velocity field $\vu$ of the cosmic fluid (left column), with the corresponding spectral ratios (right column). The five environments are indicated on the color scale on the right side.}
\label{fig:TNG50_envwps_z0}
\end{figure*}

According to the general theory of Kolmogorov turbulence, the turbulence energy spectrum can be divided into the energy-containing range, the inertial range and the dissipation range, using two characteristic scales - the integral scale and the dissipation scale \citep[e.g.][]{Davidson2015}. In our current work, we define the peak scale $k_{\rm S-peak}$ as the integral scale, and thus consider the scale range $k < k_{\rm S-peak}$ as the energy-containing range. We consider the Nyquist wavenumber $k_{\rm Nyquist}$ as the dissipation scale, which is explained below. As aforementioned, we use the PCS scheme to assign all the particles into a $N^3_{\rm g}$ grid. However, from another perspective, this assignment process is also a smoothing process -- it smooths out the velocity field at scales below the grid scale $\Delta_{\rm g}=L_{\rm box}/N_{\rm g}$, thereby eliminating turbulence at these smaller scales than $\Delta_{\rm g}$ while retaining turbulence at larger scales. In this sense, $\Delta_{\rm g}$ is the scale at which turbulence transitions into turbulence-free motion, a characteristic that is precisely indicative of the dissipation scale. Therefore, it is reasonable to consider the Nyquist wavenumber, $k_{\rm Nyquist}$, as a better representative of the dissipation scale\footnote{The Nyquist wavenumber, $k_{\rm Nyquist} \equiv \pi/\Delta_{\rm g}$, is $23.5$, $64.3$ and $137.8\hmpc$ for TNG300, TNG100 and TNG50, respectively.}. Consequently, the scale range $k_{\rm S-peak} < k < k_{\rm Nyquist}$ can be regarded as the inertial range, and the scale range $k> k_{\rm Nyquist}$ as the dissipation range of turbulence.

Note that, taking into account the numerical viscosity, the effective Reynolds number of the numerically computed flow scales as $\sim$$(L_{\rm c}/\Delta_{\rm g})^{4/3}$, where $L_{\rm c}$ is some correlation length, say the integral scale \citep{Schmidt2015}. Due to the smaller $\Delta_{\rm g}$ and therefore larger effective Reynolds number for TNG50, the TNG50 simulation is less affected by numerical viscosity, and is more suitable for turbulence studies. In the following, we will use only TNG50 data for the subsequent work.

However, it should be noted that near $k_{\rm Nyquist}$, numerical effects such as smearing, aliasing, and shot noise may be significant. Therefore, the results in this scale range should be approached with some caution \citep{Wang2024}.

Figure~\ref{fig:TNG50_velocity_globalwps_z0_4} shows the $z$-evolution of both the global energy spectrum and the spectral ratio $\tilde{Q}$ for the TNG50 simulation. We observe that the spectrum grows slightly from $z=4$ to $2$, but increases significantly from $z=2$ to $1$ in the $k < k_{\rm S-peak}$ scale range, indicating that turbulence induced by structure formation is increasingly injected into the cosmic fluid with time. However, the spectrum remains almost unchanged from $z=1$ to $0$, indicating that the turbulence is close to saturation, or that the injection of turbulence by structure formation is balanced by the transfer from the larger to the smaller scales and the dissipation of turbulence into heat. In the $k > k_{\rm S-peak}$ range, the energy spectrum is also enhanced from $z=4$ to $2$, but the magnitude of the enhancement is less than for $k < k_{\rm S-peak}$.

From $z=4$ to $z=0$, the peak scale of the spectrum increases from $k_{\rm S-peak}=27.7\hmpc$ to $k_{\rm S-peak}=8.8\hmpc$, and the transition scale from the compressive mode to the solenoidal mode also increases from $k_{\rm trans}=14.2\hmpc$ to $k_{\rm trans}=2.7\hmpc$, while the peak scale of $\tilde{Q}$ decreases from $k_{\rm Q-peak} = 27.7\hmpc$ to $k_{\rm Q-peak} = 34.9\hmpc$. Additionally, it can be seen that the exponent of the spectra at small scales becomes less steep as the redshift decreases.

In Figure~\ref{fig:TNG50_globalkurt_z0_xyz}, we show the global kurtosis as a function of $k$ for the $z=0$ TNG50 data. We can see that the three directions give almost the same results, so we will only use the $z$-direction data in the following calculation. For the $\vu$ field, it can be seen that the $\rm {kurtosis}=3$ at about $k=0.2\hmpc$, then grows rapidly with $k$ to about $k_{\rm trans}=2.7\hmpc$, and then grows slowly to the large-$k$ end, indicating that the cosmic fluid becomes increasingly intermittent as $k$ increases. As can be seen from the results for $\vu_\bot$ and $\vu_\parallel$, the kurtosis for $\vu_\bot$ is generally greater than that for $\vu_\parallel$. From the energy spectra of $\vu_\bot$ and $\vu_\parallel$ in Figure~\ref{fig:TNG_velocity_globalwps}, we know that $\vu_\parallel$ predominates in the spectra when $k<k_{\rm trans}$, but the kurtosis of $\vu_\bot$ is relatively significant. Consequently, the overall kurtosis can be considered a compromise between the contributions from $\vu_\bot$ and $\vu_\parallel$. This compromise accounts for the rapid increase in the global kurtosis observed in $k<k_{\rm trans}$. In the range $k>k_{\rm trans}$, $\vu_\bot$ predominates in the spectra, and hence the global kurtosis follows that of $\vu_\bot$, explaining the slow growth of the global kurtosis in this scale range.

In Figure~\ref{fig:TNG50_globalkurt_z0_4_z}, we show the $z$-evolution of the global kurtosis as a function of $k$ for the TNG50 data. One can see that for all four redshifts the kurtosis increases with $k$, just as for $z=0$, first in a fast mode, then in a slow growth mode. As previously analyzed, the transition scales $k_{\rm trans} = 14.2,\ 7.9,\ 5.0$ and $2.7\hmpc$ for $z=4, 2, 1$ and $0$ respectively, roughly corresponding to scales where the fast growing mode switches to the slow mode. 

At large scales of $k<3\hmpc$ the kurtosis increases monotonically with decreasing $z$, whereas at small scales of $k>3\hmpc$ the kurtosis does not change monotonically with decreasing redshift - from $z=4$ to $2$ the global kurtosis first increases and then decreases at $z<2$. These results show how the intermittency of turbulence for cosmic fluid evolves with redshift.

\begin{figure*}
\centerline{\includegraphics[width=0.975\textwidth]{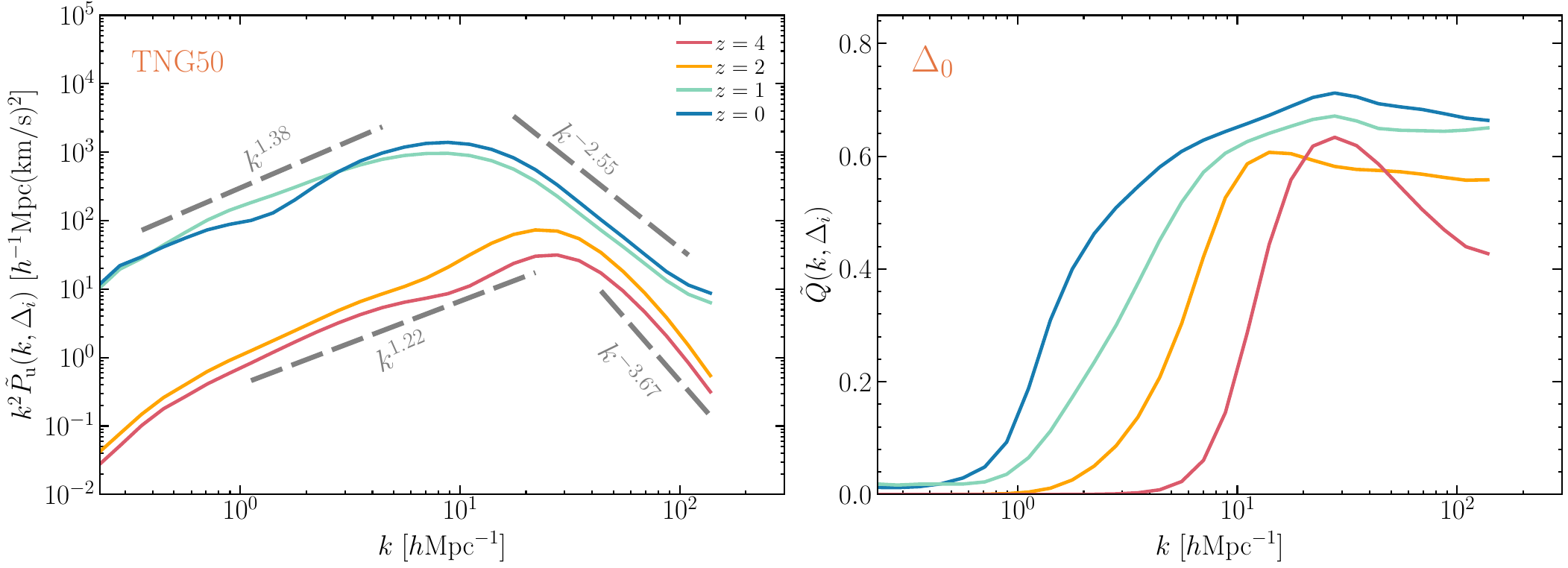}}
\centerline{\includegraphics[width=0.975\textwidth]{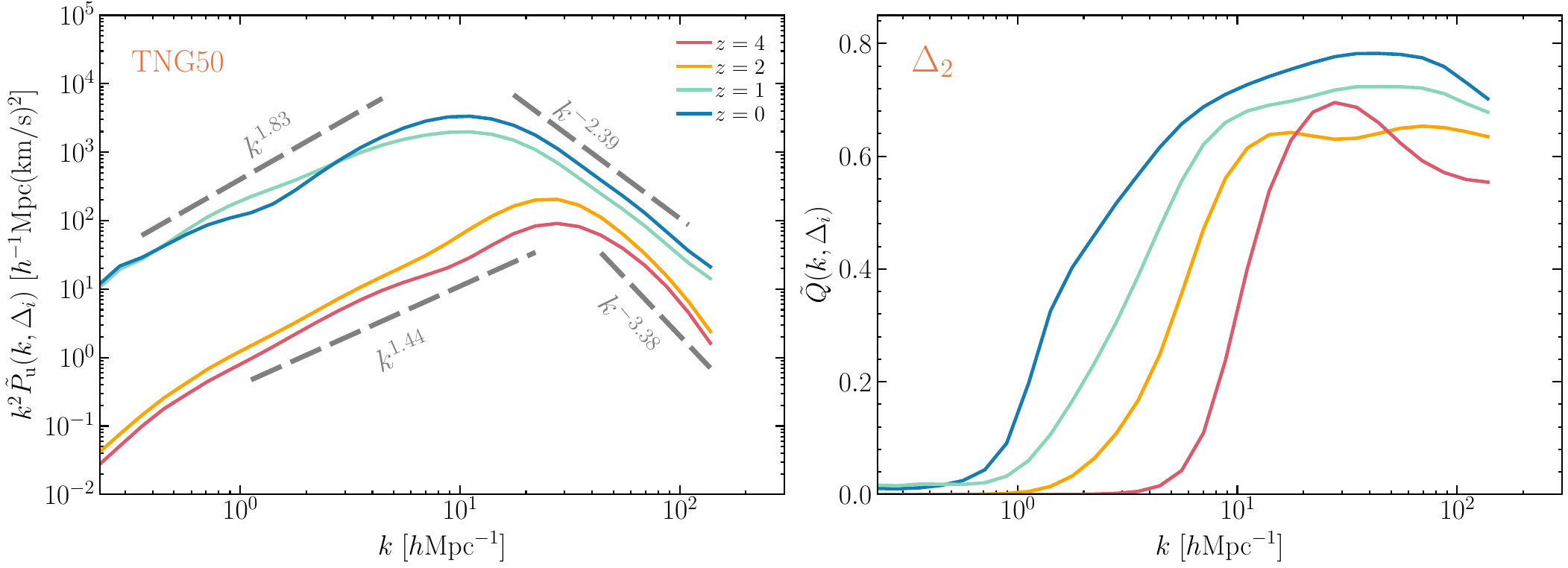}}
\centerline{\includegraphics[width=0.975\textwidth]{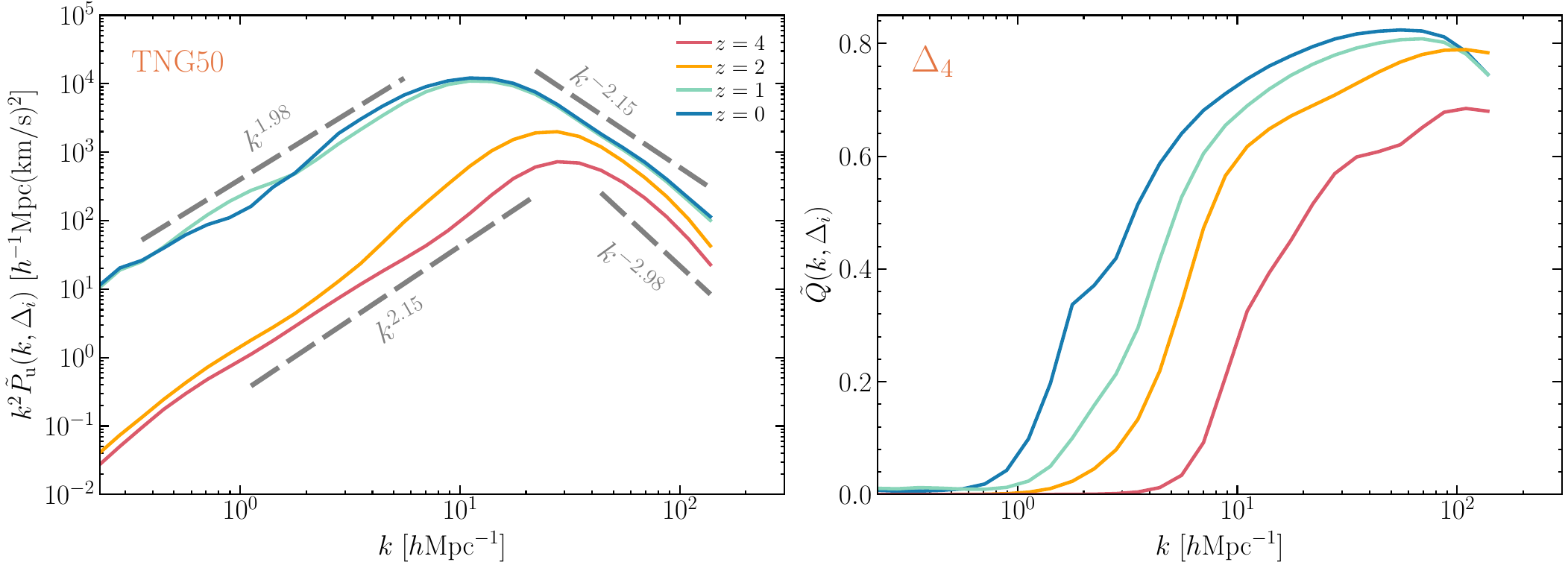}}
\caption{$z$-evolution of the energy spectrum (left column) and the spectral ratio (right column) for the TNG50 simulation of the three environments $\Delta_0$, $\Delta_2$ and $\Delta_4$, respectively. The four redshifts are shown in the figure. As in Figure~\ref{fig:TNG_velocity_globalwps}, the relevant power-laws are indicated in the figure.}
\label{fig:TNG50_envwps_Delta}
\end{figure*}

\begin{table}
\centering
\caption{The local density environments, specified with $\Delta_\mathrm{dm}=\rho_\mathrm{dm}/\bar\rho_\mathrm{dm}$ of dark matter.}
\begin{tabular}{lccccc}
\toprule
    & $\Delta_0$ & $\Delta_1$ & $\Delta_2$ & $\Delta_3$ & $\Delta_4$  \\
\hline
$\Delta_\mathrm{dm}\in$ & $[0, \ 1/8)$ & $[1/8, \ 1/2)$ & $[1/2, \ 2)$  & $[2, \ 8)$ & $[8, \ +\infty)$  \\
\hline
\end{tabular}
\label{tab:dens_envs}
\end{table}

\begin{figure}
\centerline{\includegraphics[width=0.5\textwidth]{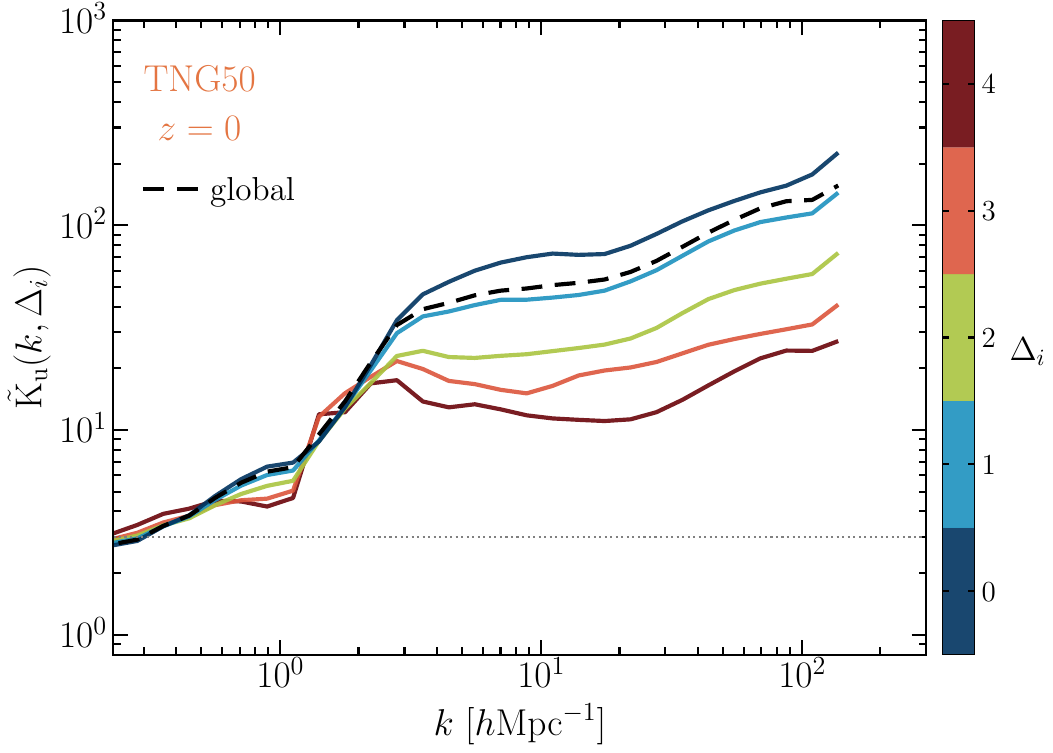}}
\caption{Environment-dependent kurtosis for the TNG50 data at $z=0$. The five environments are indicated on the color scale on the right side. The horizontal dotted line shows the ${\rm kurtosis} = 3$.}
\label{fig:TNG50_kurt_z0_z}
\end{figure}

\begin{figure*}
\centerline{\includegraphics[width=0.975\textwidth]{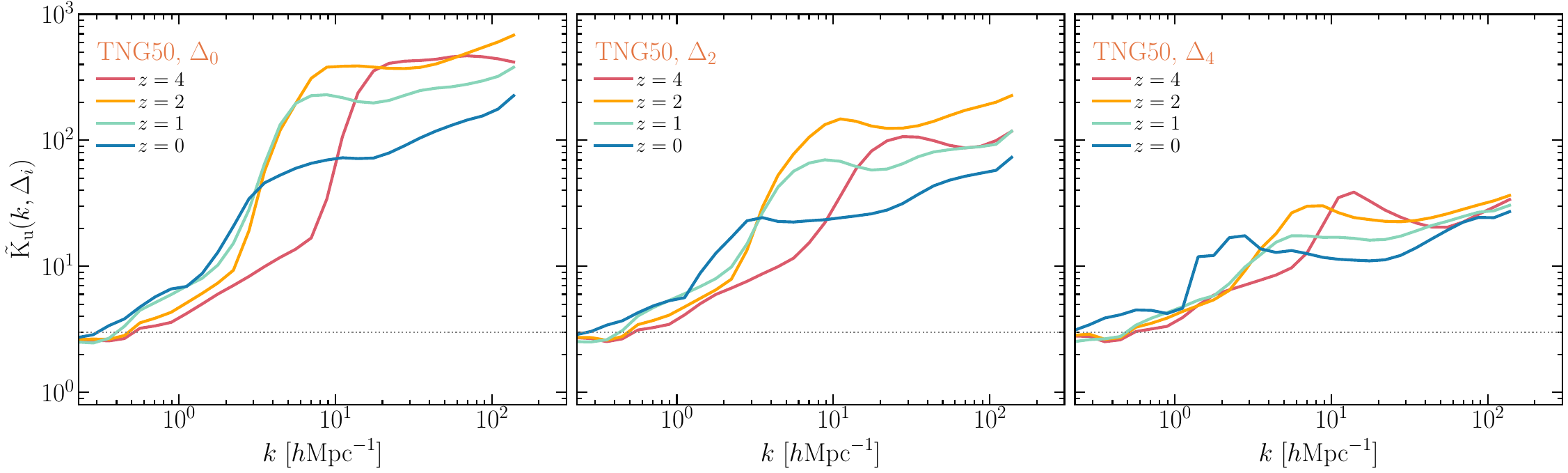}}
\caption{$z$-evolution of the env-kurtosis for the TNG50 simulation of the three environments $\Delta_0$, $\Delta_2$ and $\Delta_4$, respectively. The horizontal dotted line shows the ${\rm kurtosis} = 3$.}
\label{fig:TNG50_envkurt_z0_4_z}
\end{figure*}

\subsection{The Environment-dependent Results}

We divide the simulation space of TNG50 into five different environments according to the dark matter density, denoted as $\Delta_i$ with $i=0,1, ..., 4$ and listed in Table~\ref{tab:dens_envs}. Among them, $\Delta_0$ and $\Delta_1$ can be regarded as voids or low-density regions, and $\Delta_4$ as various high-density structures, such as clusters, filaments, sheets and their outskirts (see the bottom-right panel of Figure~\ref{fig:density_fields}). 

In Figure~\ref{fig:TNG50_envwps_z0}, we show the $z=0$ energy spectra of the velocity field $\vu$ for the five environments of TNG50. We observe that, similar to the global energy spectra in Figure~\ref{fig:TNG_velocity_globalwps}, the env-dependent spectra also exhibit peaks. The positions of these peaks are $k_{\rm S-peak} = 8.5, 9.5, 10.4, 11.1$ and $11.8\hmpc$ with increasing environmental density, respectively, indicating a trend for $k_{\rm S-peak}$ to shift towards larger values of $k$. These five env-dependent energy spectra, starting from the same $k$, increase with the increasing $k$, reach their peaks, and then decrease rapidly. These spectra are regularly distributed - the spectrum with the lowest density at the bottom and the spectrum with the highest density at the top. These results suggest that the cosmic fluid is more turbulent in a high-density environment than in a low-density environment. We also observe that the exponents of all the spectra are steeper than the Burgers exponent. Notably, the exponent in the low-density environment is even steeper than in the high-density environment, suggesting a more efficient energy transfer in low-density environment.

From the env-dependent $\tilde{Q}(k, \Delta_i)$, we see that the transition scale $k_{\rm trans} \simeq 2.7\hmpc$ is nearly identical for the low-density environments $\Delta_0$, $\Delta_1$, $\Delta_2$ and $\Delta_3$. This suggests that $\tilde{Q}(k, \Delta_i)$ is insensitive to the environmental density in the scale range $k<3\hmpc$ when $\Delta_{\rm dm}<8$. However, $k_{\rm trans}$ does depend on environmental density. For instance, in the highest-density environment $\Delta_4$, $k_{\rm trans} = 3.4\hmpc$, corresponding to $\sim1.8\mpch$, roughly the outer region within a galaxy cluster. When $k>20\hmpc$, $\tilde{Q}$ becomes sensitive to environmental density, which increases from low- to high-density environments, implying that the solenoidal mode of the velocity field is more significant in high-density environments than in low-density environments. Similar to the global spectral ratio $\tilde{Q}(k)$ shown in Figure~\ref{fig:TNG_velocity_globalwps}, the env-dependent ratios  $\tilde{Q}(k, \Delta_i)$ also exhibit peaks. The peak positions, $k_{\rm Q-peak} = 27.7, 32.2, 42.5, 51.0$ and $55.5\hmpc$, respectively, shifting to higher values with increasing environmental density. 

We observe that the energy spectra for the five density environments are quite similar, differing only in the peak position and the amplitude of the spectra. Turbulence can emerge in a dark matter void for the following reasons: (1) Even in a dark matter void, there still contains a large amount of baryonic matter, as can be seen from the baryon fraction verse dark matter density in Figure-7 of \citet{Yang2020}. (2) Mathematically, when Reynolds number is sufficiently large, say ${\rm Re} > 5000$, the solutions of the hydrodynamical equations will be turbulent. Since the viscosity in the cosmic baryonic fluid is very low, TNG employs the moving-mesh code AREPO to solve the cosmological, gravo-magnetohydrodynamical equations, disregarding viscosity. Hence, it is not difficult for a flow with low viscosity develops into turbulence even in low density regions.

In Figure~\ref{fig:TNG50_envwps_Delta}, we show the $z$-evolution of the env-dependent energy spectra and the spectral ratios $\tilde{Q}(k,\Delta_i)$ for TNG50 data from the three environments $\Delta_0$, $\Delta_2$ and $\Delta_4$ respectively. In $k < k_{\rm S-peak}$, all env-dependent spectra grow slightly from $z=4$ to $2$, but are significantly enhanced from $z=2$ to $1$, and the spectra remain almost unchanged from $z=1$ to $0$. In $k > k_{\rm S-peak}$, these energy spectra are also enhanced from $z=2$ to $1$, but the enhancement is less than for $k < k_{\rm S-peak}$. From $z=4$ to $z=0$, both the peak scales of the spectra and the transition scales increase, while the peak scales of $\tilde{Q}(k, \Delta_i)$ decrease. It can be seen that, for all the environments, the transition scales increase as the redshift decreases. Also, the exponents of all the spectra at small scales become less steep with decreasing redshift. In general, the $z$-evolution of the env-dependent energy spectra and spectral ratios is similar to the $z$-evolution of the global energy spectra and spectral ratios in Figure~\ref{fig:TNG50_velocity_globalwps_z0_4}.

In Figure~\ref{fig:TNG50_kurt_z0_z}, we show the env-kurtosis for $z=0$ TNG50 data. It can be seen that these env-dependent results are all similar to the global kurtosis - a fast growing mode followed by a slow growing mode roughly at the transition scale $k_{\rm trans}$ as $k$ increases. However, the kurtosis is largest for the lowest density environment $\Delta_0$ and decreases towards the highest density environment $\Delta_4$. The results indicate that the velocity field's intermittency in $k>k_{\rm trans}$ becomes weaker with increasing environmental density. This trend suggests an inverse relationship between the level of intermittency and the proportion of the solenoidal mode in the energy spectra, as observed by comparing the spectral ratios in Figure~\ref{fig:TNG50_envwps_z0}. In Figure~\ref{fig:TNG50_envkurt_z0_4_z}, we show the $z$-evolution of the env-kurtosis for the three environments $\Delta_0$, $\Delta_2$ and $\Delta_4$. In general, these results are similar to the $z$-evolution of the global kurtosis shown in Figure~\ref{fig:TNG50_globalkurt_z0_4_z}, except for the cases of $\Delta_0$ and $\Delta_4$, where the kurtosis does not change monotonically from $z=4$ to $2$ at small scales of $k>20\hmpc$.

These results show how the intermittency of turbulence for cosmic fluid  depends on the environmental density and evolves with redshift. In general, higher kurtosis indicates a probability distribution with a sharper peak and/or heavier tails, suggesting that there are more intense events occurring within the turbulence. Intermittent events in turbulence are often associated with localized bursts of energy release, which can be identified through changes in kurtosis. Studying intermittency in cosmic fluid can have implications for cosmological models, particularly those related to the large-scale structure of the universe and the formation of cosmic structures.

\section{Summary and Conclusions}
\label{sec:concl}

The turbulent motion of the cosmic baryonic fluid, an important topic in the study of the large-scale structures of the universe, has attracted increasing attention in cosmological studies over the last few decades. In this study, we first use CWT techniques to construct the global and environment-dependent wavelet energy spectra of the cosmic baryonic fluid, which are used to characterize the intensity of the turbulence as a function of the wavenumber $k$. The presence of shocks leads to a strong intermittency in the baryonic fluid, so that the turbulent energy dissipation in space is also intermittent. The deviation of the tails of the PDF from Gaussian is usually considered as a manifestation of intermittency. We then define the wavelet global and environment-dependent kurtosis (or flatness) of the velocity field to characterize the intermittency of turbulence in the cosmic fluid.

We use the velocity field data from the IllustrisTNG simulation. The velocity of the baryonic fluid is decomposed into turbulent and bulk flow using the iterative multiscale filtering approach of \citet{Vazza2012}, and the turbulent flow is further separated into compressive and solenoidal modes using the Helmholtz-Hodge decomposition.

We define five environments according to the dark matter density. We then compute both the global and environment-dependent energy spectrum and kurtosis of the turbulent velocity for $z=0$ TNG data, and present the $z$-evolution of the turbulent energy spectrum and kurtosis. To study the contribution of the solenoidal (or compressive) mode to the total turbulence energy, we also define the spectral ratio, either the Fourier $Q$ or the wavelet $\tilde{Q}$, of the two modes of the velocity field. Our main findings and conclusions are as follows:
\begin{enumerate}
  \item There are peaks in all the energy spectra at $k_{\rm S-peak}$ and in the spectral ratios $Q(k)$ and $\tilde{Q}(k)$ at $k_{\rm Q-peak}$ for the velocity field of the cosmic baryonic fluid. The peak scale $k_{\rm S-peak}$ can be treated as the integral scale, and the Nyquist wavenumber as the dissipation scale. Consequently, the scale range $k < k_{\rm S-peak}$, $k_{\rm S-peak} < k < k_{\rm Nyquist}$, and $k> k_{\rm Nyquist}$ can be considered as the energy-containing range, the inertial range, and the dissipation range of the turbulence, respectively. In the energy-containing range, the turbulent velocity is mostly dominated by its compressive component. The gravitational collapse converts the potential energy of the structure into kinetic energy of the bulk flow, and further decays into turbulent flows, initially in the compressive mode and, as $k$ increases, mostly in the solenoidal mode. In the inertial range, the energy passes from large-scale to small-scale eddies by the solenoidal mode or is dissipated directly into thermal energy by the compressive mode. The exponent of the energy spectrum is steeper than not only the Kolmogorov exponent but also the Burgers exponent, indicating more efficient energy transfer compared to Kolmogorov or Burgers turbulence.
  \item The global energy spectrum increases significantly from $z=2$ to $1$ in the energy-containing range, indicating that structure formation-induced turbulence is increasingly injected into the cosmic fluid with time. However, the spectrum remains almost unchanged from $z=1$ to $0$, indicating that the injection of turbulence by structure formation is balanced by the transfer from the larger to the smaller scales and the dissipation of turbulence into heat. In the inertial range, the energy spectrum is also enhanced from $z=2$ to $1$, but the magnitude of the enhancement is less than for the energy-containing range.

  \item Similar to the global energy spectra, the environment-dependent spectra can also be classified into the energy-containing, the inertial and the dissipation range based on the two characteristic scale, $k_{\rm S-peak}$ and $k_{\rm Nyquist}$. The spectral magnitude is lowest in the lowest-density environment and highest in the highest-density environment, suggesting that the cosmic fluid is more turbulent in a high-density environment than in a low-density environment. In $k>20\hmpc$, $\tilde{Q}$ increases from low- to high-density environments, implying that the solenoidal mode of the velocity field is more significant in high-density environments than in low-density environments. The $z$-evolution of the environment-dependent energy spectra and spectral ratios is generally similar to that of the global energy spectra and spectral ratios.

  \item The kurtosis (or flatness) of the cosmic velocity field is an appropriate statistic to characterize the intermittency of turbulence in the cosmic fluid. A characteristic scale $k_{\rm trans}$ can be determined by $\tilde{Q}(k_{\rm trans})$ or $Q(k_{\rm trans}) = 1/2$. For the velocity field, the wavelet global kurtosis grows rapidly with $k$ in the $k < k_{\rm trans}$ range, and then grows slowly in the $k > k_{\rm trans}$ range, indicating that the cosmic fluid becomes increasingly intermittent as $k$ increases. The kurtosis for the solenoidal component is generally greater than that the compressive component. As $z$ decreases, the kurtosis increases in the $k<3\hmpc$ range, whereas in $k>3\hmpc$ the global kurtosis first increases from $z=4$ to $2$ and then decreases at $z<2$.

  \item The environment-dependent kurtosis is almost completely similar to the global one - a fast growing mode followed by a slow growing mode at the transition scale $k_{\rm trans}$ as $k$ increases. The kurtosis is lowest for the densest environment and increases towards the lowest-density environment, indicating that the intermittency of the velocity field in the $k>k_{\rm trans}$ range becomes increasingly strong as the density of the environments decreases. The $z$-evolution of the env-kurtosis is similar to that of the global kurtosis, except for the cases of the lowest- and highest-density environment, where the kurtosis does not change monotonically from $z=4$ to $2$ in the $k>20\hmpc$ range. 
\end{enumerate}

Comparisons of the global WPS and the spectral ratios $\tilde{Q}(k)$ with the corresponding results of the Fourier analysis show that the Fourier analysis and the global wavelet analysis give almost the same results, demonstrating that the wavelet analysis techniques are reliable and trustworthy. However, the environment-dependent wavelet analysis proves to be more powerful than the Fourier analysis in that the wavelet statistics such as the env-WPS or the env-spectral ratios $\tilde{Q}(k, \Delta_i)$ cannot be implemented by the Fourier analysis. In this work, the environments are simply defined by the dark matter density. In fact, one can design various complex environments by specifying, for example, a particular spatial region or a specific cosmic structure.

The simple theoretical framework of turbulence is based on Kolmogorov turbulence, which is homogeneous and isotropic in space and is characterized by eddies of different scales. However, this picture is problematic for the turbulence of the cosmic baryonic fluid, which is subject to structure formation driven by gravity in the context of cosmic expansion. Structure formation leads, for example, to the density stratification of the cosmic fluid, where buoyancy forces resist radial motions, making the turbulence anisotropic \citep[e.g.][]{Shi2018, Shi2019, Mohapatra2020, Simonte2022, Wangc2023}. In this work, we have not yet considered the anisotropic turbulence caused by the density stratification, which will be left for future studies.

Following the general theory of Kolmogorov turbulence, we also divide the turbulent energy spectrum of the cosmic baryonic fluid into three ranges, the energy-containing range, the inertial range and the dissipation range. Note that these authors \citep{Schmidt2009, Schmidt2010, Schmidt2016, Shi2018, Shi2019, Shi2020} also refer to these terms, but they should adopt different definitions from ours. Indeed, as \citet{Shi2018} points out, these concepts are not easy to identify and define. In fact, the real situation is much more complex than this general picture. Since the compressive mode exists at all scales, the kinetic energy of the turbulent flow, even of the bulk flow, can be directly dissipated into thermal energy at all scales. 

In addition, in the inertial and dissipation range, both kinetic and thermal energy may also be injected by feedback from SN-driven galactic winds or AGN outflows, and these physical processes deserve to be studied in detail. Thus, the fact that the spectral exponent in the inertial range is steeper than that of Kolmogorov or Burgers turbulence, indicates that the simple theory of Kolmogorov or Burgers turbulence does not hold for the cosmic baryonic fluid. One can speculate that a steeper energy spectrum may be of great significance for structure formation, which, for example, can lead to more efficient transfer of energy to smaller scales, potentially affecting the formation and evolution of galaxies.

\section*{Acknowledgments}

The authors thank the anonymous referee for helpful comments and suggestions. We acknowledge the use of the data from IllustrisTNG simulation for this work. We also acknowledge the support by the National Science Foundation of China (No. 12147217, 12347163), and by the Natural Science Foundation of Jilin Province, China (No. 20180101228JC).

\software{NumPy\footnote{\url{https://numpy.org/}}\citep{vander2011,Harris2020}, nbodykit\footnote{\url{https://nbodykit.readthedocs.io/en/latest/}}\citep{Hand2018}, SciPy\footnote{\url{https://scipy.org/}}\citep{Jones2001}, Matplotlib\footnote{\url{https://matplotlib.org/}}\citep{Hunter2007}, WPSmesh\footnote{\url{https://github.com/WangYun1995/WPSmesh}}\citep{Wang2024}, Jupyter Notebook\footnote{\url{https://jupyter.org/}}.}

\appendix
\restartappendixnumbering

\section{The match between FPS and global WPS}
\label{sec:fps_vs_wps}

There are a wide variety of wavelets, e.g. Poisson wavelet, Morse wavelet, Morlet wavelet, and b-spline wavelet, as well as CW-GDW we use here. The fact that they have different forms implies, on the one hand, that the same wavelet scale value may represent different physical scales for different wavelets, and on the other hand, the WPS computed from different wavelets has different magnitudes. Therefore, we need to standardize the wavelet scale and the magnitude of WPS, which can be done by matching them to the wavenumber and FPS, respectively.

Following the scheme of \citet{Meyers1993} and \citet{Torrence1998}, we have demonstrated in Sec 2.5 of \citet{Wang2022a} that the relationship $w=c_w k$ between the wavelet scale $w$ and wavenumber $k$ can be obtained by solving $\partial|W_{\cos}(w,x)|^2/\partial w=0$ for $w$, where $W_{\cos}(w,x)=\cos (kx)\hat\psi(k,w)$ is the CWT of the 1D cosine function $\cos (k x)$. With a closer look, we see that $\partial|W_{\cos}(w,x)|^2/\partial w=0$ is equivalent to $\partial|\hat\psi(w,k)|^2/\partial w=0$, and further to
\begin{equation}
\partial|\hat\psi(c_w,1)|^2/\partial c_w=0.
\label{eq:cw_eqn}
\end{equation}
In accordance with Equation \eqref{eq:cw_eqn}, we can get the value $c_w\approx 0.3883$ for the isotropic CW-GDW by solving $\partial|\hat\Psi(c_w,1)|^2/\partial c_w=0$ numerically.

Next, integrating both sides of Equation \eqref{eq:gwps_fps} with respect to $w$ from zero to infinity, we have
\begin{align}
    & \int_0^{+\infty }\tilde P_\mathbf{u}(w)\mathrm{d}w = \nonumber\\
    & \quad \frac{1}{2\pi^2}\int_0^{+\infty}P_\mathbf{u}(k) \left(\int_0^{+\infty}|\hat\Psi(w,k)|^2\mathrm{d}w\right)k^2\mathrm{d}k.
    \label{eq:fps_vs_wps_1}
\end{align}
Let $k'=k/w$, then we have
\begin{equation}
    \int_0^{+\infty}|\hat\Psi(w,k)|^2\mathrm{d}w =\frac{1}{k^2}\int_0^{+\infty}k'|\hat\Psi(k')|^2\mathrm{d}k', \nonumber
\end{equation}
which can be substituted into the Equation \eqref{eq:fps_vs_wps_1} to get
\begin{equation}
    \int_0^{+\infty }\tilde P_\mathbf{u}(w)\mathrm{d}w=\frac{I_\Psi}{2\pi^2}\int_0^{+\infty}P_\mathbf{u}(k) \mathrm{d}k,\nonumber 
\end{equation}
where $I_\Psi=\int_0^{+\infty}k|\hat\Psi(k)|^2\mathrm{d}k$ is a constant depending on the wavelet. Using the $w=c_wk$ and $\tilde P_\mathbf{u}(k)\equiv\tilde P_\mathbf{u}(c_wk)$, we further obtain
\begin{equation}
    \int_0^{+\infty }\tilde P_\mathbf{u}(k)\mathrm{d}k=\int_0^{+\infty}\frac{I_\Psi}{2\pi^2c_w}P_\mathbf{u}(k) \mathrm{d}k.\nonumber
\end{equation}
Considering the similar shapes of $\tilde P_\mathbf{u}(k)$ and $P_\mathbf{u}(k)$, the above equation implies that the integrand terms on both sides are approximately equal, i.e.
\begin{equation}
    \tilde P_\mathbf{u}(k)\approx\frac{I_\Psi}{2\pi^2c_w}P_\mathbf{u}(k),
    \label{eq:fps_vs_wps_2}
\end{equation}
in which the value of $I_\Psi/(2\pi^2c_w)$ is roughly 1.0754 for the isotropic CW-GDW.

Note that the above procedure does not assume a particular form of the power spectrum. In fact, we can get a more accurate conclusion for the power-law spectra, $P_\mathbf{u}(k)=Ak^n$. Substituting it into Equation \eqref{eq:gwps_fps} gives

\begin{table*}[t]
\centering
\caption{The values of scaling factor $c_w^nI_n/2\pi^2$ at different indices.}
\begin{tabular}{lccccccc}
\toprule
 $n$   & $-3$ & $-2$ & $-1$ & $0$ & $1$ & $2$ & $3$ \\
\hline 
$c_w^nI_n/2\pi^2$ & $1.4738$ & $1.2211$ & $1.0745$  & $1$ & $0.9763$ & $0.9963$ & $1.0587$\\ [+0.2em]
\hline
\end{tabular}
\label{tab:wps_factor}
\end{table*}

\begin{equation}
    \tilde P_\mathbf{u}(k) =\frac{c_w^n}{2\pi^2}I_nP_\mathbf{u}(k),
    \label{eq:fps_vs_wps_3}
\end{equation}
where $I_n=\int_0^{+\infty}k^{n+2}|\hat\Psi(k)|^2\mathrm{d}k$ is a function of index $n$. For clarity, we list the values of $c_w^nI_n/2\pi^2$ for different spectral indices in Table \ref{tab:wps_factor}. As can be seen, Equation \eqref{eq:fps_vs_wps_2} holds exactly only for $n=-1$. However, most power spectra in the real world are more complicated than a simple power-law form, and it is not necessary or practicable to utilize a different factor value for each segment with a different index. Therefore, it is a fair approximation to match the FPS by multiplying the entire WPS by a factor $I_\Psi/(2\pi^2c_w)$, at least for $n>-3$.

\section{Relationship between Exponent of Energy Spectrum and Energy Transfer Rate}
\label{sec:relation_exp_etr}

\begin{figure}
\centerline{\includegraphics[width=0.495\textwidth]{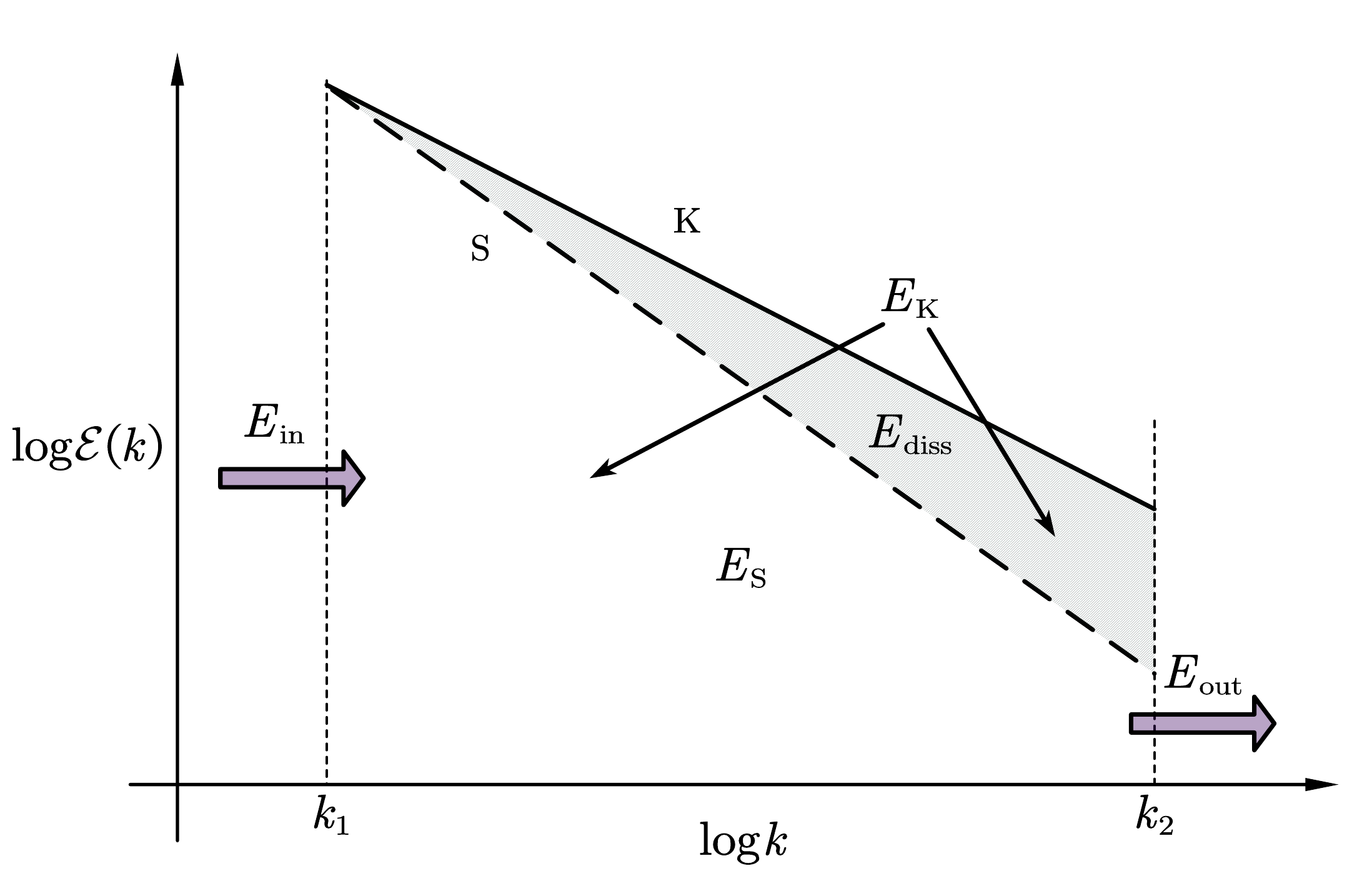}}
\caption{An illustration of the relationship between the energy spectrum's exponent and the rate of energy transfer. `K’ denotes the spectrum for Kolmogorov turbulence, and `S’ denotes the energy spectrum for the turbulence with a steeper exponent. $k_1$ and $k_2$ correspond to the integral scale $l_{\rm int}$ and dissipation scale $l_{\rm diss}$, respectively, with $k_1=2\pi/l_{\rm int}$ and  $k_2=2\pi/l_{\rm diss}$. $E_{\rm in}$ denotes the injection energy, $E_{\rm K}$ denotes the turbulent energy of Kolmogorov turbulence, $E_{\rm S}$ denotes the turbulent energy of steeper-energy-spectrum turbulence, $E_{\rm diss}$ denotes the energy that is dissipated in the inertial range, and $E_{\rm out}$ denotes the energy that is dissipated at the dissipation scale $k_2$. In the following analysis, we assume that the shapes of all energy spectra, such as the exponents and the amplitudes of the spectra, have been balanced by the continuously injected energy, thus remaining unchanged over time.}
\label{fig:espect}
\end{figure}

For homogeneous and isotropic turbulence, its energy spectrum in the inertial range can usually be expressed in the form of a power law as $\propto k^{-n}$, such as $n = 5/3, 2$ to represent Kolmogorov and Burgers turbulence, respectively. A general conclusion is: A steeper turbulence energy spectrum, with $n>5/3$, has a higher rate of energy transfer from turbulent kinetic energy to thermal energy. We will elucidate this point with Figure~\ref{fig:espect} below.

For a turbulent fluid system, we denote its turbulent kinetic energy as $E_{\rm turb}$, and its energy transfer rate from turbulent kinetic energy to thermal energy at the dissipation scale $k_2$ as $\varepsilon$. This energy transfer rate is a characteristic of the turbulent fluid system and should not depend on, for example, the exponent of the energy spectrum. Hence, the time it takes to transfer the turbulent kinetic energy from large-scale eddies to small-scale eddies and then dissipate it into thermal energy at the dissipation scale $k_2$ is given by
\begin{align}
\label{eq:trans_time}
\Delta t = \frac{E_{\rm out}}{\varepsilon},
\end{align}
where $E_{\rm out}$ is actually the total turbulent energy within the inertial range, i.e. $E_{\rm out} = E_{\rm turb}$. For a turbulent fluid, its turbulent energy within the inertial range is given by:
\begin{align}
\label{eq:turb_energy}
E_{\rm turb} = \int_{k_1}^{k_2}\mathcal{E}(k) \dd k,
\end{align}
where $\mathcal{E}(k)$ is the energy spectrum of the turbulence.

Since there is no energy dissipation within the inertial range for Kolmogorov turbulence, the energy transfer rate for Kolmogorov turbulence is:
\begin{align}
\label{eq:K_epsilon}
\varepsilon_{\rm K} = \frac{E_{\rm in}}{\Delta t} = \frac{E_{\rm K}}{E_{\rm out}/\varepsilon} = \varepsilon,
\end{align}
where we use Equation~(\ref{eq:trans_time}) and $E_{\rm in} = E_{\rm K} = E_{\rm out}$. $E_{\rm K}$ is the turbulent energy of Kolmogorov turbulence, derived with Equation~(\ref{eq:turb_energy}).

Now let's turn to the case of the steeper-spectrum turbulence `S', whose turbulent energy within the inertial range is denoted as $E_{\rm S}$, calculated with Equation~(\ref{eq:turb_energy}). From Figure~\ref{fig:espect}, we know that
\begin{align}
\label{eq:S_relation}
E_{\rm in} = E_{\rm S} + E_{\rm diss},
\end{align}
where $E_{\rm diss}$ is the energy that is NOT transferred from large-scale to small-scale eddies, but is directly dissipated into thermal energy through shock heating within the inertial range. Such processes occur within very thin fronts of shock-waves and should be considered to occur instantaneously, i.e. the time taken is negligible. It should be noted that the time required to transfer energy into heat is not determined by the injection energy $E_{\rm in}$, but by the turbulent kinetic energy $E_{\rm out} (= E_{\rm S})$, as suggested by Equation~\ref{eq:trans_time}. Hence the energy transfer rate for the steeper-spectrum turbulence is
\begin{align}
\label{eq:S_epsilon}
\varepsilon_{\rm S} & = \frac{E_{\rm in}}{\Delta t} = \frac{E_{\rm S} + E_{\rm diss}}{E_{\rm out}/\varepsilon} = \frac{E_{\rm S} + E_{\rm diss}}{E_{\rm S}/\varepsilon} \nonumber \\
& = (1+\frac{E_{\rm diss}}{E_{\rm S}})\varepsilon > \varepsilon = \varepsilon_{\rm K},
\end{align}
where we use Equations~(\ref{eq:trans_time}), (\ref{eq:K_epsilon}) and (\ref{eq:S_relation}). From Equation~(\ref{eq:S_epsilon}), we see that the energy transfer rate for the steeper-spectrum turbulence is larger than that for Kolmogorov turbulence.

This is precisely our conclusion that for homogeneous and isotropic turbulence, a steeper energy spectrum of turbulence indicates a more efficient energy transfer.

\bibliography{references}{}
\bibliographystyle{aasjournal}

\end{CJK*}
\end{document}